\documentclass[sigconf]{acmart}

\usepackage{tablefootnote}
\usepackage{amsmath,amsfonts}
\usepackage{algorithmic}
\usepackage{graphicx}
\usepackage{xcolor}
\usepackage{hyperref}
\usepackage{setspace}
\usepackage{booktabs} 
\usepackage{subfigure}
\usepackage{amsmath}
\usepackage{color}
\usepackage{amsmath}
\usepackage{mathrsfs}
\usepackage[normalem]{ulem}
\usepackage{enumitem}
\usepackage{multirow}
\usepackage{ulem}
\usepackage[nomargin,inline,marginclue,draft]{fixme}
\usepackage{balance}
\usepackage{verbatim}
\usepackage{diagbox}
\usepackage{changepage}
\usepackage{xcolor}
\usepackage{booktabs}
\usepackage{bbm}
\usepackage{adjustbox}

\usepackage{subcaption}

\usepackage{footnote}

\usepackage[normalem]{ulem}
\useunder{\uline}{\ul}{}

 \copyrightyear{2024}
\acmYear{2024}
\setcopyright{rightsretained}
\acmConference[SIGIR '24]{Proceedings of the 47th International ACM SIGIR Conference on Research and Development in Information Retrieval}{July 14--18, 2024}{Washington, DC, USA}
\acmBooktitle{Proceedings of the 47th International ACM SIGIR Conference on Research and Development in Information Retrieval (SIGIR '24), July 14--18, 2024, Washington, DC, USA}
\acmDOI{10.1145/3626772.3657802} \acmISBN{979-8-4007-0431-4/24/07}

\makeatletter
\gdef\@copyrightpermission{
  \begin{minipage}{0.3\columnwidth}
   \href{https://creativecommons.org/licenses/by/4.0/}{\includegraphics[width=0.90\textwidth]{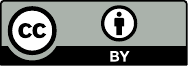}}
  \end{minipage}\hfill
  \begin{minipage}{0.7\columnwidth}
   \href{https://creativecommons.org/licenses/by/4.0/}{This work is licensed under a Creative Commons Attribution International 4.0 License.}
  \end{minipage}
  \vspace{5pt}
}
\makeatother

\begin{document}
\makesavenoteenv{itemize}

\title{Modeling User Fatigue for Sequential Recommendation}

\author{Nian Li}
\orcid{0000-0003-4689-2289}
\authornote{Contribute equally to this work.}
\affiliation{
  \institution{Shenzhen International Graduate School, Tsinghua University}
  \city{Shenzhen}
  \country{China}
}

\author{Xin Ban}
\authornotemark[1]
\author{Cheng Ling}
\affiliation{%
  \institution{Kuaishou Inc.}
  \city{Beijing}
  \country{China}
}

\author{Chen Gao}
\authornote{Corresponding author (chgao96@gmail.com, liyong07@tsinghua.edu.cn).}
\affiliation{%
  \institution{Tsinghua University}
  \city{Beijing}
  \country{China}
}

\author{Lantao Hu}
\author{Peng Jiang}
\affiliation{%
  \institution{Kuaishou Inc.}
  \city{Beijing}
  \country{China}
}

\author{Kun Gai}
\affiliation{%
  \institution{Independent}
  \city{Beijing}
  \country{China}
}

\author{Yong Li}
\authornotemark[2]
\affiliation{%
  \institution{Tsinghua University}
  \city{Beijing}
  \country{China}
}

\author{Qingmin Liao}
\affiliation{
  \institution{Shenzhen International Graduate School, Tsinghua University}
  \city{Shenzhen}
  \country{China}
}

\renewcommand{\shortauthors}{Nian Li et al.}

\begin{abstract}

Recommender systems filter out information that meets user interests. However, users may be tired of the recommendations that are too similar to the content they have been exposed to in a short historical period, which is the so-called \textit{user fatigue}. 
Despite the significance for a better user experience, user fatigue is seldom explored by existing recommenders.
In fact, there are three main challenges to be addressed for modeling user fatigue, including what features support it, how it influences user interests, and how its explicit signals are obtained.
In this paper, we propose to model user \textbf{F}atigue in interest learning for sequential \textbf{Rec}ommendations~(\textbf{FRec}). To address the first challenge, based on a multi-interest framework, we connect the target item with historical items and construct an interest-aware similarity matrix as features to support fatigue modeling. Regarding the second challenge, built upon feature cross, we propose a fatigue-enhanced multi-interest fusion to capture long-term interest. In addition, we develop a fatigue-gated recurrent unit for short-term interest learning, with temporal fatigue representations as important inputs for constructing update and reset gates. For the last challenge, we propose a novel sequence augmentation to obtain explicit fatigue signals for contrastive learning. We conduct extensive experiments on real-world datasets, including two public datasets and one large-scale industrial dataset. Experimental results show that FRec can improve AUC and GAUC up to 0.026 and 0.019 compared with state-of-the-art models, respectively. Moreover, large-scale online experiments demonstrate the effectiveness of FRec for fatigue reduction. Our codes are released at \url{https://github.com/tsinghua-fib-lab/SIGIR24-FRec}.

\end{abstract}


\maketitle

\section{Introduction}
In today's online platforms, the recommender system is broadly deployed to filter out irrelevant content and fetch personalized content for users that they are interested in~\cite{hamilton2017inductive,ding2019reinforced,ding2020simplify,quan2023robust,gao2024causal}. Therefore, in the development of recommendation models, how to capture user interests as accurately as possible is an essential problem. 

Sequential recommender organizes users' historical interactions in a temporal sequence and aims to predict the next item of interaction~\cite{chang2021sequential,zheng2022disentangling}. Many existing works built upon advanced neural networks focus on interest learning, including long and short-term user interests~\cite{hidasi2015session,yu2019adaptive,zheng2022disentangling,tang2018personalized}. Some works also propose to combine long and short-term interest modeling for better recommendation~\cite{an2019neural,yu2019adaptive,zheng2022disentangling}. Another line of modeling accurate user interests is to extract multiple interests from the sequence~\cite{pi2019practice,lian2021multi,cen2020controllable}. These works argue that only one representation for modeling interests is not effective enough since users are usually interested in several kinds of items.

\begin{figure}[t!]
    \centering
    \subfigure[Illustration demo]{
    \includegraphics[width=0.23\textwidth]{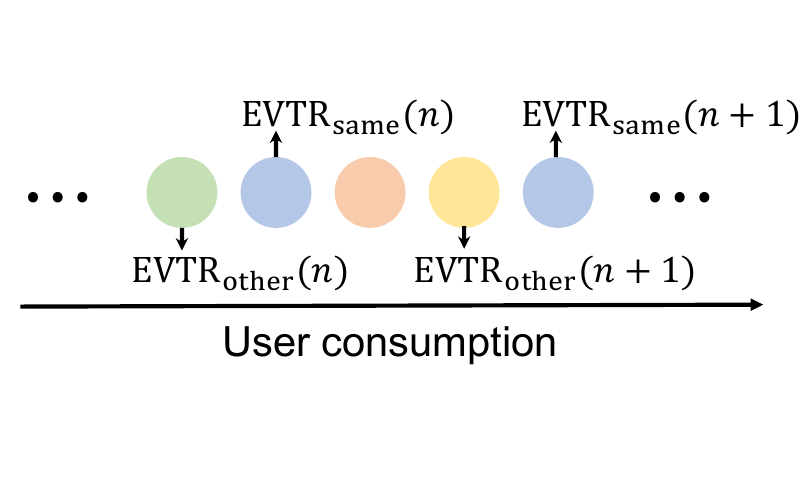}}
    \subfigure[EVTR trend]{
    \includegraphics[width=0.23\textwidth]{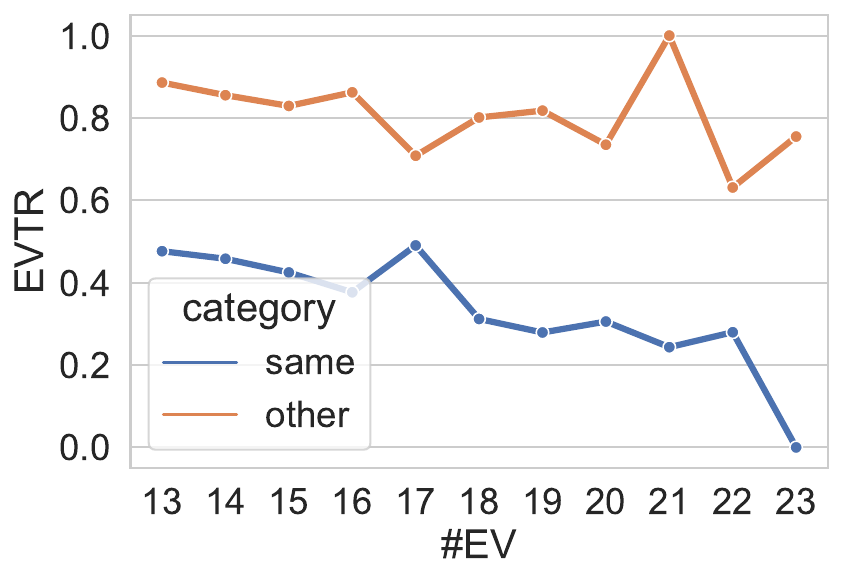}}
    \caption{(a) An illustration demo to show how we measure EVTR along with user consumption. Color represents the video category, and blue is the target category to count the number of effective views. (b) The trend of EVTR with increasing consumption of videos with the same category. We include the EVTR of the video with the target category~(same) and videos with other categories close to it~(other).}
   \label{fig:evtr}
\end{figure}

Despite of this, user fatigue has not been well studied in existing works, especially how it can influence user interests. In this work, \textbf{user fatigue refers that users may be tired of the recommendations that are too similar to content they have been exposed to in a short historical period}, such as news, advertisements, \textit{etc.} For example, the click-through rate~(CTR) of news will drop significantly with more and more times of exposure~\cite{xie2022multi}. It is important to note that user fatigue is fundamentally different from other concepts related to positive user experience in recommender systems. Specifically, \textit{diversity} typically focuses solely on the dissimilarity between items in the recommendation list, irrespective of the user's historical interactions~\cite{alhijawi2022survey}. On the other hand, \textit{serendipity} or \textit{novelty} emphasizes that the recommended items are unexpected or unknown to the user, characterized by their divergence from historical items or by the items' popularity~\cite{fu2023deep}. In contrast, user fatigue represents the negative aspect of user experience.
We verify the existence of user fatigue on a micro-video platform Kuaishou, using large-scale interaction data involving tens of millions of users. An industrial dataset is also collected from this platform for experiments in Section \ref{exp}. In Figure \ref{fig:evtr}, we plot the \textit{normalized} effective view-through rate~(EVTR) as the function of the number of effective views of videos with the same category in historical consumption. Compared with videos with other categories, the EVTR of videos with the target category decreases significantly and is consistently lower when users have too many effective views of the same category. This is obvious evidence of user fatigue with respect to the repetitive consumption of similar videos. This issue can harm user experience and further reduce platform activity.

A few existing works address the issue of user fatigue with coarse-grained features based on item-level and category-level repetitions. Ma \textit{et al.}~\cite{ma2016user} just feed these features into decision trees, which serve as the base recommendation model. Moriwaki \textit{et al.}~\cite{moriwaki2019fatigue} define a simple quadratic function for directly mapping the features to user fatigue. These methods are usually ineffective since the way of modeling fatigue lacks flexibility and interpretability. As a matter of fact, there are three challenges to be addressed,
\begin{itemize}[leftmargin=*]
    \item \textbf{Fine-grained features are hard to obtain to support fatigue modeling.} Intuitively, user fatigue depends on the similarity between the target and historical items. Existing works usually utilize item-level and category-level repetitions as the similarity features~\cite{ma2016user, xie2022multi}. However, these measurements are usually too coarse to represent the similarity between items accurately. For instance, even if the two videos both belong to the category of `pandas', there may be still non-negligible differences, such as one is about `panda is eating bamboo' and the other is about `pandas rentals from the UK'. Therefore, how to measure fine-grained similarity to support fatigue modeling is critical but difficult.
    \item \textbf{The influence of user fatigue on interests is complex.} In general, the user's certain interest will be weakened if he/she is experiencing fatigue with it. Existing works either neglect to model this influence~\cite{ma2016user, xie2022multi} or manually define it by a quadratic function~\cite{moriwaki2019fatigue}, which is unrealistic in real-world scenarios. Actually,  multiple historical items may contribute to causing user fatigue as a whole and further influence both long and short-term interests. Therefore, based on similarity features, how to fuse user fatigue with interest learning is also an essential point.
    \item \textbf{There are no explicit signals of user fatigue contained in historical consumption.} 
    The decreasing engagement with certain types of items over time can be seen as users are tired of frequent exposures. However, this phenomenon can only be observed from later consumption after the current interaction. Therefore, it is hard to directly obtain corresponding signals of user fatigue with respect to the current item from historical consumption.
\end{itemize}

In this work, we propose to model user fatigue in interest learning for sequential recommendations with the challenges above addressed. Specifically, we first extract multi-interest representations\footnote{We use ``representation'' and ``embedding'' interchangeably in this paper.} from the historical sequence with the self-attention mechanism. To obtain fine-grained features to support fatigue modeling, we construct an interest-aware similarity matrix~(ISM) measured by the projection distance built upon historical and target item embeddings. We then apply cross networks for feature interplay to assist in handling complex fatigue influence, based on which we model the influence on long-term interest. We further develop a fatigue-gated recurrent unit~(FRU) for short-term interest learning. 
For explicit signals of user fatigue, we propose a novel sequence augmentation to obtain them counterfactually and use them to supervise contrastive learning with respect to fatigue prediction.

We have conducted extensive experiments on two public datasets and one large-scale industrial dataset to evaluate the effectiveness of our FRec. Compared with many state-of-the-art~(SOTA) models, our FRec achieves significant improvements with respect to various accuracy and ranking metrics. Further online studies also demonstrate that FRec can reduce user fatigue by alleviating repeated exposure in consecutive consumptions and improve user experience significantly.

The contributions of this work are summarized as follows,
\begin{itemize}[leftmargin=*]
    \item We take user fatigue into consideration and make an advanced step to incorporate it into interest learning for sequential recommendations.
    \item We address primary challenges in modeling user fatigue by constructing fine-grained similarity features, handling its complex influence on long and short-term interests, and obtaining its signals with a novel sequence augmentation for contrastive learning.
    \item We conduct extensive offline and online experiments to demonstrate that FRec can improve the recommendation accuracy significantly (AUC up to  0.026, GAUC up to 0.019, and NDCG up to 5.8\%) compared with SOTA methods and reduce user fatigue. 
\end{itemize}

\section{Problem Formulation}
We consider a standard problem of sequential recommendation. For each user $u\in \mathcal{U}$, let $S_u=\{i_1,i_2,\cdots,i_{L_u}\}$ denote the historical interaction sequence chronologically, \textit{i.e.}, ordered by the interaction timestamp of each item, where $i_l\in\mathcal{I}$ and $L_u$ is the sequence length. $\mathcal{U}$ and $\mathcal{I}$ denote the set of users and items, respectively.

Most existing works focus on modeling user interests according to historical sequence $S_u$ for predicting whether the user will interact with the target item $i_t$. However, user fatigue is also a critical factor influencing the user interest and decision. In other words, if the item $i_t$ is very similar to many items in $S_u$, the user may not interact with it due to being tired of repeated interactions. In this work, we aim to model user fatigue with the sequence $S_u$ and incorporate it with interest learning for capturing user decisions more accurately.\\
\noindent\textbf{Input}: Historical sequences for all the users $\{S_u\vert u\in\mathcal{U}\}$.\\
\textbf{Output}: A model that can predict the user's interaction probability for the next~(target) item $i_t$.

\section{Method}
\begin{figure}[t!]
    \centering
    \includegraphics[scale=0.38]{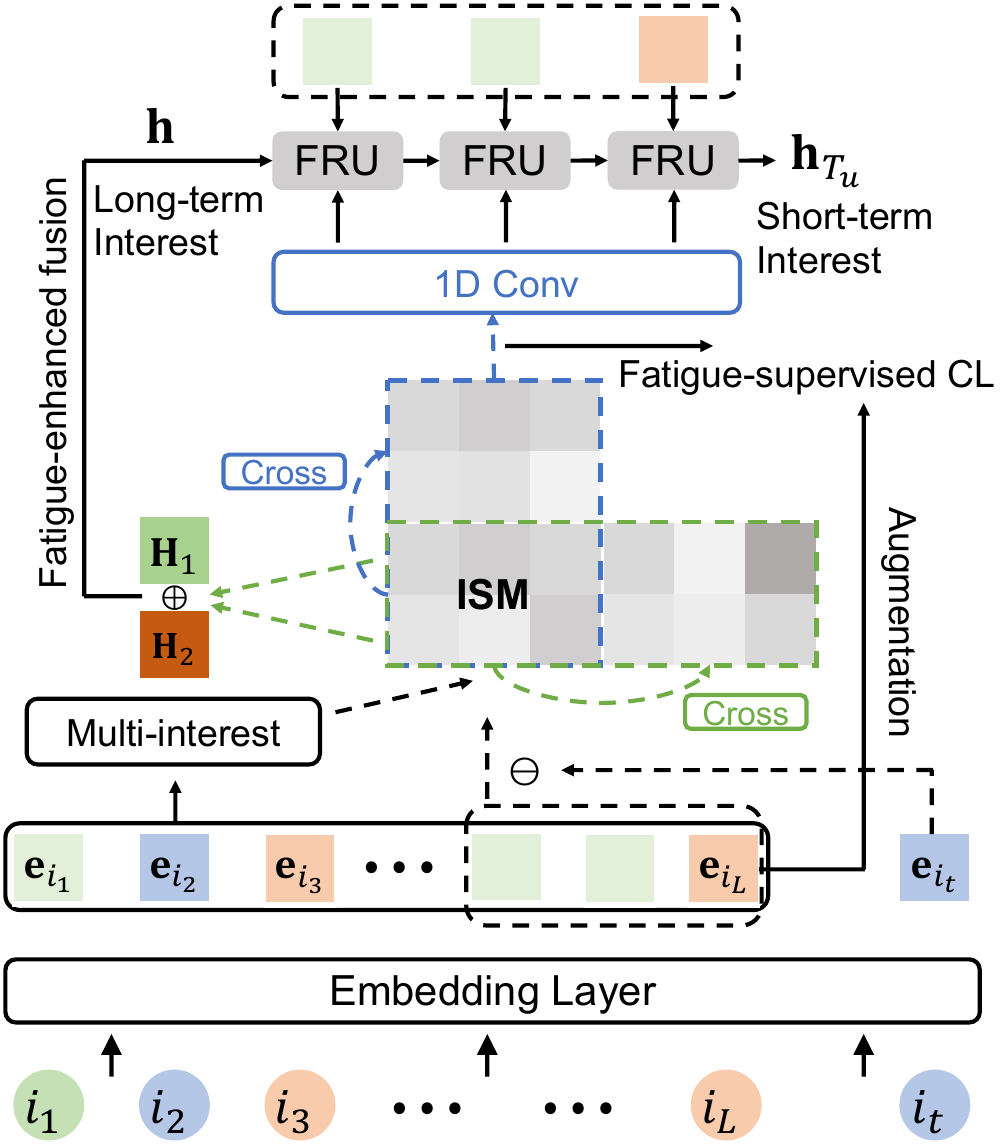}
    \caption{The framework of FRec.}
    \label{fig:frec}
\end{figure}

Figure \ref{fig:frec} shows the framework of our FRec with four modules.

\begin{table}[t!]
\caption{Frequently used notations.}
    \centering
    \begin{tabular}{c|l}
    \toprule
        $S_u$, $L_u$ & Historical sequence for user $u$, and its length. \\
        $\hat{S}_u$, $T_u$ & Sub-sequence with recent items in $S_u$, and its length. \\
        $T$ & Truncated threshold for selecting sub-sequence $\hat{S}_u$. \\
        $K$ & The number of interests. \\
        $C$ & The number of cross and convolutional layers. \\
        \midrule
        $\mathbf{M}_j (\mathbf{M}^\top_j)$ & The $j$-th column of the matrix $\mathbf{M}$~(transposed $\mathbf{M}^\top$). \\
        $\mathbf{e}_{i}$ & The embedding of item $i$. \\
        $\mathbf{H}$ & Multi-interest embedding matrix. \\
        $\mathbf{F}$ & Interest-aware similarity matrix. \\
        $\mathbf{h}, \mathbf{h}_{T_u}$ & Long and short-term interest embedding. \\
        $\mathbf{MLP}$ & Multi-layer perceptron applied on the last dimension. \\
        $\mathbf{W}, \mathbf{b}$ & Learnable weight matrix and bias vector. \\
    \bottomrule
    \end{tabular}
    \label{tab:notation}
\end{table}

\subsection{Interest-aware Similarity Matrix}
Fine-grained target-historical item similarity is necessary to support the modeling of user fatigue. Indeed, the similarity between two items can stem from multiple aspects and correspond to multiple sub-interests of the user. For instance, the similarity of videos can be characterized by aspects such as shooting style, video tone, and topics, all of which can be used to model user fatigue when watching videos. Therefore, we first extract multi-interests from historical sequences.

First of all, each item $i$ is assigned an embedding $\mathbf{e}_i\in\mathbb{R}^{d\times 1}$, where $d$ is the embedding dimension. Correspondingly, the sequence $S_u$ for user $u$ can be encoded as an embedding matrix $\mathbf{S}_u\in \mathbb{R}^{d\times L_u}$, where the $l$-th column is $\mathbf{e}_{i_l}$, the embedding for item $i_l$ in the sequence. We then choose a widely-used self-attention mechanism~\cite{lin2017structured,cen2020controllable} for multi-interest extraction. Specifically, we generate a multi-interest embedding matrix for user $u$ as follows,
\begin{equation}
\begin{aligned}
    &\mathbf{H} = \mathbf{S}_u\mathbf{A}, \\
    &\mathbf{A} = \mathit{softmax}(\mathbf{MLP}_1(\mathbf{S}_u^\top)),
\end{aligned}
\end{equation}
where $\mathbf{MLP}_1$ is a two-layer perceptron with \textit{tanh} as nonlinear activation, and the output dimension is the number of interests $K$, which is a tunable hyper-parameter. Here $\mathbf{A}\in \mathbb{R}^{L_u\times K}$ is attention weights for aggregating all the item embeddings in the sequence, which is generated by applying $\mathit{softmax}$ along with the first dimension of $\mathbf{MLP}_1$ output. Finally, we obtain $K$ interest embeddings $\mathbf{H}\in \mathbb{R}^{d\times K}$ from the user's historical interactions.

\begin{figure}
    \centering
    \includegraphics[scale=0.42]{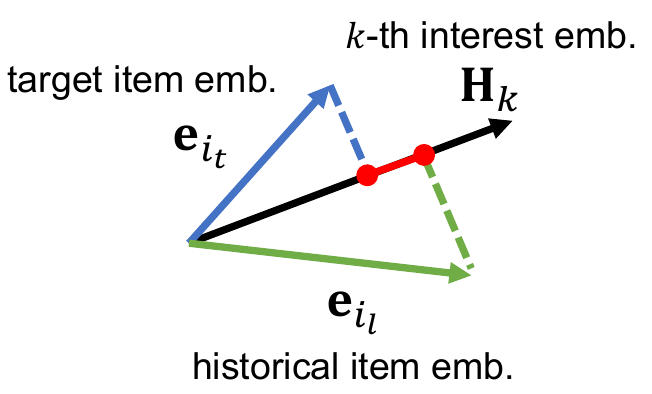}
    \caption{Illustration of the calculation of interest-aware similarity, represented by the length of the red line.}
    \label{fig:sim}
\end{figure}

To obtain fine-grained target-historical similarity, we leverage extracted multi-interest and item embeddings in latent space. Compared with existing works utilizing coarse-grained item-level or category-level features~\cite{ma2016user,xie2022multi}, the embedding-based similarity can measure relevance more accurately and effectively. Specifically, we construct an interest-aware similarity matrix to measure the similarity between the target item $i_t$ and historical item $i_l$ with respect to each user interest, formulated as follows,
\begin{equation}
    \mathbf{F}_{l,k} = \frac{1}{1+\left\vert\frac{\mathbf{e}_{i_t}^\top \mathbf{H}_k}{\Vert\mathbf{H}_k\Vert}-\frac{\mathbf{e}_{i_l}^\top \mathbf{H}_k}{\Vert\mathbf{H}_k\Vert}\right\vert},
\end{equation}
where the similarity is based on the projection distance between the embedding of $i_t$ and $i_l$ on the $k$-th interest embedding $\mathbf{H}_k$, and shorter distance means higher similarity. Figure \ref{fig:sim} illustrates how this similarity is calculated. Considering that user fatigue is the most relevant to items nearest to the target item, we confine the calculation of this similarity feature among the most recent $T_u=\mathrm{min}(T,L_u)$ items, \textit{i.e.}, $l\in \{L_u-T_u+1, L_u-T_u+2, \cdots, L_u\}$. $T$ is a tunable truncated threshold to control how many recent items should be included. We denote this sub-sequence of items as $\hat{S}_u$. The similarity matrix $\mathbf{F}\in\mathbb{R}^{T\times K}$ will be padded with zeros as $\mathbf{F}^\top = [\mathbf{F}^\top, \mathbf{0}^\top_{K\times (T-L_u)}]$ if $T>L_u$, where $[\cdot, \cdot]$ denotes the concatenation operation along the last dimension. It will support the modeling of user fatigue along with capturing users' long and short-term interests.

\subsection{Fatigue-enhanced Multi-interest Fusion}
Although multi-interest embeddings $\mathbf{H}$ capture multiple aspects of interests, there is a critical influence of user fatigue on long-term interests with respect to the target item. In other words, the importance should be decreased if the user is experiencing fatigue for a certain sub-interest. To adaptively adjust long-term interest, we propose a fatigue-enhanced multi-interest fusion built upon interest-aware similarity matrix $\mathbf{F}$. A direct way is,
\begin{equation}
\label{eqn:fusion}
\begin{aligned}
    &\mathbf{h} = \mathbf{H}\mathbf{w}, \\
    &\mathbf{w} = \mathit{softmax}(\mathbf{MLP}_2(\mathbf{F}^\top)),
\end{aligned}
\end{equation}
where $\mathbf{MLP}_2$ has the output dimension of 1, and attention weights $\mathbf{w}\in\mathbb{R}^{K\times 1}$ for interests fusion are obtained from similarity features with respect to each sub-interest. However, there are two key difficulties when learning fatigue-aware importance. On the one hand, the dependency of user fatigue on target-historical similarity can be nonlinear or even more complex~\cite{aharon2019soft,moriwaki2019fatigue}, and directly feeding these features into neural networks may not guarantee accurate modeling. On the other hand, several recently consumed items can jointly contribute to user fatigue on the target item. For example, compared with consuming only one video, a user experience much more fatigue if he/she has consumed five videos published by the same author of the target item within one hour. To tackle these problems, inspired by \cite{wang2017deep}, we propose to utilize feature cross and apply $C$ layers of Cross Network on the similarity matrix. Specifically, each cross layer processes the $c$-th features as follows,
\begin{equation}
    \mathbf{P}_{c+1} = \mathbf{P}_0\odot (\mathbf{W}_c\mathbf{P}_c) + \mathbf{P}_c,
\end{equation}
where $\mathbf{W}_c\in\mathbb{R}^{T\times T}$ is a learnable weight matrix and $\odot$ denotes element-wise product. The layer $c$ ranges from 0 to $C-1$ and $\mathbf{P}_0=\mathbf{F}$. In this way, the feature interplay of the same item can generate high-order features for modeling complex similarity-fatigue dependency. The interplay between different items can assist in modeling the effect of multiple items on user fatigue. Finally, Eq. \ref{eqn:fusion} can be modified as follows,
\begin{equation}
\begin{aligned}
    &\mathbf{h} = \mathbf{H}\mathbf{w}, \\
    &\mathbf{w} = \mathit{softmax}\left(\mathbf{MLP}_2\left([\mathbf{P}_C^\top, \mathbf{P}_0^\top]\right)\right).
\end{aligned}
\end{equation}
With the fatigue-enhanced fusion, we obtain the user's long-term interest embedding $\mathbf{h}\in \mathbb{R}^{d\times 1}$.

\subsection{Fatigue-gated Recurrent Unit}
\label{fru}
As stated in the previous subsection, recently consumed items in the historical sequence are important in causing user fatigue. Therefore, we model the influence of temporal user fatigue on short-term interest learning. Similarly, the similarity features are first processed with cross networks to tackle the problems of complex similarity-fatigue dependency and joint effects from multiple interests. Specifically, $C$ cross layers are applied as follows,
\begin{equation}
    \mathbf{Q}_{c+1} = \mathbf{Q}_0\odot (\mathbf{Q}_c\mathbf{W}'_c) + \mathbf{Q}_c,
\end{equation}
where $\mathbf{W}'_c\in\mathbb{R}^{K\times K}$ is a learnable weight matrix and $\mathbf{Q}_0=\mathbf{F}$. Furthermore, the temporal pattern contained in the sequence of recent items is also necessary for modeling fatigue. For instance, consuming five consecutive items similar to the target item will cause a more heightened perception of fatigue than that of disordered ones. Inspired by the effectiveness of CNNs in modeling sequences~\cite{tang2018personalized,bai2018empirical}, we apply 1D convolutional networks to further model temporal user fatigue sequentially. Each layer of convolution operation for the $l$-th item in the sub-sequence $\hat{S}_u$ is formulated as follows,
\begin{equation}
\label{eqn:conv}
\begin{aligned}
    &\hat{\mathbf{Q}}_l^\top = [q_l^1, q_l^2, \cdots, q_l^{d_{\mathrm{out}}}]^\top, \\
    &q_l^n = \mathit{LeakyRelu}\left(\mathrm{SUM}\left(\hat{\mathbf{Q}}_{l-s+1:l}^\top\odot \mathbf{W}_{\mathrm{conv}}^n\right)\right),
\end{aligned}
\end{equation}
where $\mathbf{W}_{\mathrm{conv}}^n\in \mathbb{R}^{d_{\mathrm{in}}\times s}$ is the $n$-th learnable filter kernel, and $d_{\mathrm{in}}$ and $s$ denotes the input dimension and the kernel size respectively. The number of filter kernels~(\textit{i.e.}, the output dimension of this convolutional layer) is $d_{\mathrm{out}}$. The input is the crossed features obtained above, \textit{i.e.}, the initial $\hat{\mathbf{Q}} = [\mathbf{Q}_C, \mathbf{Q}_0]$, thus the initial $d_{\mathrm{in}} = 2K$. After $C$ layers of convolution, we model temporal fatigue until the $l$-th item in the representation $\hat{\mathbf{Q}}_l$. Note that we use `causal' convolutions~\cite{bai2018empirical} since current fatigue only depends on previous items. Zero padding is utilized when $l < s$.

In terms of modeling short-term interest, RNNs have been demonstrated as effective modules in many advanced works~\cite{hidasi2015session,yu2019adaptive,zhou2019deep,zheng2022disentangling}, such as GRU, LSTM, \textit{etc.} To incorporate fatigue influence, we propose a fatigue-gated recurrent unit~(FRU) built upon GRU. Specifically, the extracted fatigue representation until each item serves as additional feature input to construct update and reset gates. With the state input $\mathbf{h}_{l-1}\in \mathbb{R}^{d_{\mathrm{in}}\times 1}$ from previous step and embedding input $\mathbf{x}_{l}\in \mathbb{R}^{d_{\mathrm{in}}\times 1}$, new state $\mathbf{h}_l\in \mathbb{R}^{d_{\mathrm{out}}\times 1}$ is calculated as follows,
\begin{equation}
\begin{aligned}
    &\mathbf{z}_l = \mathit{sigmoid}(\mathbf{W}_z \mathbf{x}_l + \mathbf{U}_z \mathbf{h}_{l-1} +  \underline{\mathbf{V}_z \hat{\mathbf{Q}}_l} + \mathbf{b}_z), \\
    &\mathbf{r}_l = \mathit{sigmoid}(\mathbf{W}_r \mathbf{x}_l + \mathbf{U}_r \mathbf{h}_{l-1} + \underline{\mathbf{V}_r \hat{\mathbf{Q}}_l} + \mathbf{b}_r), \\
    &\hat{\mathbf{h}}_l = \mathit{tanh}\left(\mathbf{W}_h \mathbf{x}_l + \mathbf{U}_h(\mathbf{r}_l \odot \mathbf{h}_{l-1}) + \mathbf{b}_h\right), \\
    &\mathbf{h}_l = (1-\mathbf{z}_l)\odot \mathbf{h}_{l-1} + \mathbf{z}_l \odot\hat{\mathbf{h}}_l,
\end{aligned}
\end{equation}
where $\mathbf{W}_{z,r,h},\mathbf{U}_{z,r,h},\mathbf{V}_{z,r}\in \mathbb{R}^{d_{\mathrm{out}}\times d_{\mathrm{in}}}$ and $\mathbf{b}_{z,r,h}\in \mathbb{R}^{d_{\mathrm{out}}\times 1}$ are learnable weights and bias. The embedding input is the $l$-th item's embedding, \textit{i.e.}, $\mathbf{x}_l = \mathbf{e}_{i_l}$. We set the initial state $\mathbf{h}_0 = \mathbf{h}$, which is the long-term interest embedding with fatigue-enhanced fusion. In this formula, temporal user fatigue affects how short-term interests evolve. Generally speaking, the interests before the current time step should not be propagated to the next step if the corresponding fatigue is intense. We apply FRU on the sub-sequence $\hat{S}_u$, and the final output $\mathbf{h}_{T_u}\in \mathbb{R}^{d_{\mathrm{out}}\times 1}$ encodes user's short-term interests with fatigue influence.

\subsection{Fatigue-supervised Contrastive Learning}
\begin{figure}
    \centering
    \includegraphics[scale=0.38]{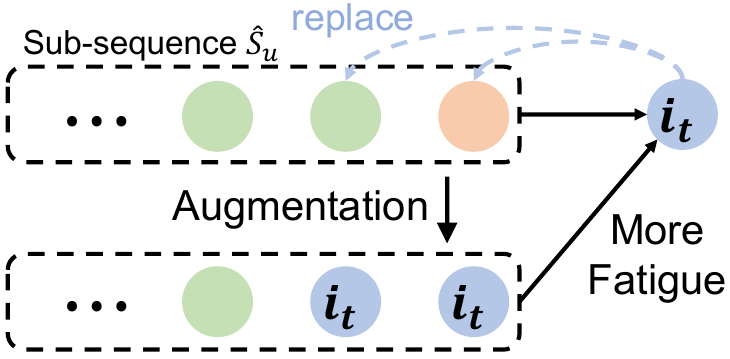}
    \caption{Illustration of the sequence augmentation to obtain fatigue signals. There is more fatigue if some items in the sub-sequence $\hat{S}_u$ are replaced by the target item.}
    \label{fig:cl}
\end{figure}

Although we have encoded temporal fatigue in latent space, whether these representations guarantee the modeling of real fatigue is unknown. The challenge is that there are no explicit signals for the supervision of representation learning. Inspired by the advantages of self-supervised learning~\cite{yu2022self} and multi-task learning~\cite{quan2023alleviating} in recommendations, we propose a novel sequence augmentation for fatigue-supervised contrastive learning. Specifically, $N\in [\mathrm{max}(N_r, 1), T_u]$ items in the sub-sequence $\hat{S}_u$ are replaced by the target item, where $N_r$ is the number of repetitions of the target item in $\hat{S}_u$. As shown in Figure \ref{fig:cl}, the primary idea is that users will experience more fatigue if they have much more repetitive consumption. We set a margin $N_r$ when choosing how many items to replace, which is large enough from the experimental results. 

With the augmented sequence, we can also obtain similarity features $\mathbf{Q}_0'$~(and the processed $\mathbf{Q}'_C$ by cross networks) motivating the modeling of temporal fatigue after the same modeling introduced in previous subsections. All the learnable parameters are shared when modeling original and augmented sequences. Finally, user fatigue for the interaction of the target item can be predicted as follows,
\begin{equation}
\begin{aligned}
    f = \mathrm{MEAN}\left(\mathbf{MLP}_3\left([\mathbf{Q}_C, \mathbf{Q}_0]\right)\right), \\
    f' = \mathrm{MEAN}\left(\mathbf{MLP}_3\left([\mathbf{Q}'_C, \mathbf{Q}'_0]\right)\right).
\end{aligned}    
\end{equation}
All the items in $\hat{S}_u$ are considered for modeling the fatigue with $\mathrm{MEAN}$.
We then formulate the contrastive loss with fatigue as the supervision as follows,
\begin{equation}
    \mathcal{L}_{\mathrm{con}} = \sum-\log \frac{\exp(-f)}{\exp(-f)+\sum_{j=1}^4\exp(-f'_j)},
\end{equation}
where $f'_j$ denotes the fatigue of the $j$-th augmentation, and we conduct four augmentations for each instance. Note that the fatigue should be larger after the augmentation, thus we use $-f$ to calculate the likelihood.

\subsection{Model Training}
We consider both users' long-term and short-term interests for interaction predictions, as well as user fatigue when making the decision on the target item. The prediction score for the user $u$ and the target items $i_t$ is calculated as follows,
\begin{equation}
    y_{u, i_t} = \mathbf{MLP}_4([\mathbf{h}^\top, \mathbf{h}_{T_u}^\top, \mathbf{e}_{i_t}^\top]) - \mathit{tanh}(f).
\end{equation}
We explicitly decrease the score with predicted user fatigue obtained from the subsection above. The function $\mathit{tanh}$ controls the magnitude of effects.

The recommendation loss for model training is a widely-used softmax loss function~\cite{cen2020controllable}, formulated as follows,
\begin{equation}
    \mathcal{L}_{\mathrm{rec}} = \sum\limits_{(u,i_t,i'_1\sim i'_4)\in\mathcal{O}}-\log \frac{\exp(y_{u,i_t})}{\exp(y_{u,i_t})+\sum_{j=1}^4\exp(y_{u,i'_j})},
\end{equation}
where $\mathcal{O}$ denotes all the training data, and $i'_1\sim i'_4$ are randomly sampled negative items for each $(u, i_t)$ pair.

Final training loss is the combination of recommendation and contrastive loss,
\begin{equation}
    \mathcal{L} = \mathcal{L}_{\mathrm{rec}} + \alpha \mathcal{L}_{\mathrm{con}},
\end{equation}
where $\alpha$ is a hyper-parameter for controlling the importance of fatigue supervision.

\section{Experiments}
\label{exp}
To evaluate our proposed FRec, we conduct extensive experiments on both public and large-scale industrial datasets. We will answer the following research questions in this section,
\begin{itemize}[leftmargin=*]
    \item \textbf{RQ1}: Can FRec outperform state-of-the-art models in terms of recommendation accuracy?
    \item \textbf{RQ2}: Can each key module benefit the overall performance?
    \item \textbf{RQ3}: Can FRec result in the reduction of user fatigue?
    \item \textbf{RQ4}: How do key hyper-parameters influence the performance?
\end{itemize}

\subsection{Experimental Settings}

\textbf{Datasets.}
The statistics of datasets are shown in Table \ref{tab:dataset}, where Avg. Length denotes the sequence length averaged over all the users. 
\begin{itemize}[leftmargin=*]
    \item \textbf{Kuaishou}\footnote{https://www.kuaishou.com}. This is one of the largest micro-video platforms in China and this dataset has been used in many related works~\cite{zheng2022disentangling,chang2021sequential}. It contains users' interactions with micro-videos over one week (October 22 to October 28, 2020), and records various behaviors such as click, like, follow, \textit{etc.} We extract the click interactions for experiments. 
    \item \textbf{Taobao}\footnote{https://tianchi.aliyun.com/dataset/dataDetail?dataId=649}. This is the largest e-commerce platform in China. This dataset records users' interactions with various products from November 25 to December 3, 2017, including page view, cart, purchase, \textit{etc.} We follow existing works~\cite{zheng2022disentangling} and choose the data of page view for experiments. 
    \item \textbf{Industrial}. The interaction data is collected from Kuaishou for 1 hour, involving tens of millions of users. Unlike public Kuaishou dataset, we include various behavioral data for experiments to model user fatigue from uninterrupted behavioral sequences.
\end{itemize}

We adopt widely-used 10-core rules~\cite{chang2021sequential,zheng2022disentangling} for public datasets to filter out inactive users and unpopular items. We split sequential interactions chronologically into 8:1:1 for training, validating, and testing models~\cite{cen2020controllable}. Since we model user fatigue with respect to both short-term and long-term interests~(such as a period of several days), the maximum sequence length is set longer than average length, \textit{i.e.}, $250$ for Kuaishou dataset and $100$ for Taobao dataset.

\begin{table}[t!]
    \centering
    \caption{Statistics of three datasets.}
    \begin{tabular}{c|c|c|c|c}
    \toprule
         Dataset & \#Users & \#Items & \#Instances & Avg. Length\\
         \midrule
         Kuaishou & 37,502 & 131,063 & 6,427,764 & 171.4\\
         Taobao & 41,101 & 90,524 & 2,256,967 &  54.9\\
         Industrial & 38,467,817 & 19,863,454 & 804,934,827 & 20.9\\
    \bottomrule
    \end{tabular}
    \label{tab:dataset}
\end{table}

\noindent\textbf{Baselines.}
We choose the following state-of-the-art~(SOTA) recommendation models for comparisons, 1) long-term and~(or) short-term interest modeling: \textbf{DIN}~\cite{zhou2018deep}, \textbf{DIEN}~\cite{zhou2019deep}, \textbf{GRU4Rec}~\cite{hidasi2015session}, \textbf{SASRec}~\cite{kang2018self}, \textbf{AdaMCT}~\cite{jiang2023adamct}, \textbf{Caser}~\cite{tang2018personalized}, \textbf{SLi-Rec}~\cite{yu2019adaptive}, and \textbf{CLSR}~\cite{zheng2022disentangling}, 2) multi-interest modeling: \textbf{SUM}~\cite{lian2021multi}, \textbf{ComiRec}~\cite{cen2020controllable} with two versions of extracting multiple interests by dynamic routing~(-DR) or self-attention~(-SA), and \textbf{MGNM}~\cite{tian2022multi}, 3) fatigue modeling\footnote{Although there is another method modeling user fatigue for click-through rate prediction~\cite{li2023fan}, we don't include it because modeling fatigue relies on \textit{non-click} historical sequences and rich context features. Besides, it obtains supervision signals of user fatigue through interaction data in future three days. These are unusual settings and not applicable to our general problem.}: \textbf{DFN}~\cite{xie2022multi}.

\noindent\textbf{Evaluation Metrics.}
Similar to existing works sampling one negative item for each positive instance~\cite{yu2019adaptive,chang2021sequential,zheng2022disentangling}, we sample nine negative items to ensure robust training and evaluation~\cite{lin2022dual}. We adopt widely-used accuracy metrics AUC and GAUC~\cite{zhou2018deep} as well as ranking metrics HR@k, NDCG@k, and MRR~\cite{zheng2022disentangling,chang2021sequential,lin2022dual,tian2022multi} for performance evaluation. We set $k$ as 2 and 4, a widely-used setting in existing works~\cite{chang2021sequential,zheng2022disentangling}.

\noindent\textbf{Hyper-parameter Settings.}
We implement our FRec and all the baselines with Microsoft Recommender\footnote{ https://github.com/microsoft/recommenders} based on Tensorflow\footnote{ https://www.tensorflow.org}. We use Adam~\cite{kingma2014adam} optimizer for modeling learning, where the initial learning rate is 0.001. L2 regularization weight is searched among \{1e-4, 1e-6\}. The batch size for training is set as 500. We early stop the training process when GAUC on the validation set decreases for two consecutive epochs. Embedding dimension $d$ is set as 40 for all the models. The $\mathbf{MLP}_4$ for final prediction is three-layer with hidden size [100, 64], with \textit{relu} as activation function, and batch normalization. We conduct a careful grid search to find optimal hyper-parameters for each model, following the original papers. For our FRec, the kernel size of convolutional layers $s=5$, the number of interests $K=4$, and the number of cross and convolutional layers $C=2$. For the convolution, the number of filter kernels~(\textit{i.e.}, hidden dimension) is [20, 40]. In FRU, $d_{in} = d_{out} = 40$. Regarding other $\mathbf{MLP}_{1,2,3}$, they are two-layer with the hidden size as half of the input dimension. The truncated threshold $T$ is set as $50$ for Kuaishou and $40$ for the Taobao dataset. The weight of contrastive learning $\alpha=0.4$. Note that these hyper-parameters are not carefully tuned for better performance.

\subsection{Overall Comparison~(RQ1)}
\textbf{Public Datasets}. The performance on the Kuaishou and Taobao datasets are shown in Table \ref{tab:pub_perf}. From the comparison, we have the following observations,
\begin{itemize}[leftmargin=*]
    \item \textbf{FRec outperforms all the baselines significantly.} On the Kuaishou dataset, FRec improves by about 0.009 in terms of AUC and GAUC. Corresponding improvements are 0.026 and 0.019 on the Taobao dataset. For other ranking metrics, the improvements range from 1.3\%$\sim$3.0\% and 2.4\%$\sim$ 5.8\% on the Kuaishou and Taobao datasets, respectively. The p-value < 0.001 demonstrates that FRec can give consistently and significantly more accurate recommendations than SOTA models.
    \item \textbf{Modeling long and short-term interests or multi-interests can obtain better performance generally.} CLSR, a SOTA model disentangling users' long and short-term interests based on causal structure, obtains almost the best performance among the baselines on both datasets. Compared with models only capturing long-term~(\textit{e.g.}, DIN) or short-term~(\textit{e.g.}, GRU4Rec) interest, jointly modeling~(CLSR, SLi-Rec) is better on most metrics. ComiRec-SA, a multi-interest framework based on self-attention, also obtains competitive performance. 
    \item \textbf{Feeding coarse-grained similarity features can benefit model performance.} On both datasets, DFN outperforms the backbone DIN on most metrics, especially on AUC. However, top performance can not be guaranteed since it directly concatenates several fatigue-aware features~(\textit{e.g.}, the number of historical consumed items, the number of items that belong to the same category with the target item) with the embedding of the target item. In other words, it's necessary to capture complex influence of user fatigue on interests accurately in model design.
\end{itemize}

\begin{table*}[t!]
\centering
\caption{Performance comparison on public dataset. All the results are averaged over five experiments. {\ul Underline} means the best two baselines, and \textbf{bold} means p-value < 0.001 compared with the best baseline under the student's t-test.}
\scalebox{0.86}{
\begin{tabular}{cc|cccccccc|cccc|cc}
\toprule
\multirow{2}{*}{}    & \multirow{2}{*}{Model} & \multirow{2}{*}{DIN} & \multirow{2}{*}{DIEN} & \multirow{2}{*}{GRU4Rec} & \multirow{2}{*}{SASRec} & \multirow{2}{*}{AdaMCT} & \multirow{2}{*}{Caser} & \multirow{2}{*}{SLi-Rec} & \multirow{2}{*}{CLSR} & \multirow{2}{*}{SUM} & \multirow{2}{*}{\shortstack{ComiRec\\-DR}} & \multirow{2}{*}{\shortstack{ComiRec\\-SA}} & \multirow{2}{*}{\shortstack{MGNM}}
 & \multirow{2}{*}{DFN} & \multirow{2}{*}{FRec} \\
 &&&&&&&&&&&&&& \\ 
\midrule
\multirow{7}{*}{\rotatebox[origin=c]{90}{\textbf{Kuaishou}}} 
& AUC & 0.6054 & 0.7520 & {\ul 0.8306} & 0.8298 & 0.8067 & 0.8228 & 0.8258 & 0.8263 & 0.8235 & 0.8239 & {\ul 0.8441} & OOM\tablefootnote{Due to the requirement of constructing graphs based on historical sequences, MGNM encounters out-of-memory~(OOM) on the Kuaishou dataset, which has a longer maximum sequence length of $250$. In original paper, the maximum sequence length is only $100$.} & 0.6613 & \textbf{0.8533} \\
&GAUC & 0.8204 & 0.8198 & 0.8401 & 0.8270 & 0.8033 & 0.8417 & 0.8388 & {\ul 0.8473} & 0.8414 & 0.8259 & {\ul 0.8464} & OOM & 0.8159 & \textbf{0.8564} \\
&HR@2 & 0.6179 & 0.6249 & 0.6570 & 0.6226 & 0.5776 & 0.6552 & 0.6651 & {\ul 0.6703} & 0.6570 & 0.6301 & {\ul 0.6658} & OOM & 0.6284 & \textbf{0.6878} \\
&HR@4 & 0.8269 & 0.8356 & 0.8642 & 0.8466 & 0.8172 & 0.8683 & 0.8585 & {\ul 0.8747} & 0.8670 & 0.8429 & {\ul 0.8705} & OOM & 0.8424 & \textbf{0.8860} \\
&NDCG@2 & 0.5417 & 0.5484 & 0.5784 & 0.5428 & 0.4982 & 0.5749 & {\ul 0.5897} & {\ul 0.5901} & 0.5779 & 0.5523 & 0.5869 & OOM & 0.5509 & \textbf{0.6077} \\
&NDCG@4 & 0.6403 & 0.6479 & 0.6765 & 0.6486 & 0.6112 & 0.6758 & 0.6812 & {\ul 0.6869} & 0.6772 & 0.6527 & {\ul 0.6837} & OOM & 0.6519 & \textbf{0.7016} \\
&MRR & 0.6045 & 0.6111 & 0.6355 & 0.6073 & 0.5719 & 0.6327 & {\ul 0.6442} & {\ul 0.6444} & 0.6353 & 0.6143 & 0.6422 & OOM & 0.6136 & \textbf{0.6583} \\
\midrule
\multirow{7}{*}{\rotatebox[origin=c]{90}{\textbf{Taobao}}} 
& AUC & 0.6800 & 0.7592 & 0.8257 & {\ul 0.8455} & 0.8412 & 0.8264 & 0.8333 & {\ul 0.8527} & 0.8247 & 0.7820 & 0.8359 & 0.7291 & 0.7630 & \textbf{0.8795} \\
&GAUC & {\ul 0.8469} & 0.8263 & 0.8327 & 0.8430 & 0.8336 & 0.8376 & 0.8381 & {\ul 0.8601} & 0.8281 & 0.7779 & 0.8333 & 0.7279 & 0.8459 & \textbf{0.8792} \\
&HR@2 & 0.7072 & 0.6737 & 0.6922 & 0.6964 & 0.6842 & 0.6878 & 0.6857 & {\ul 0.7305} & 0.6818 & 0.5675 & 0.6667 & 0.4897 & {\ul 0.7144} & \textbf{0.7660} \\
&HR@4 & {\ul 0.8585} & 0.8393 & 0.8331 & 0.8460 & 0.8325 & 0.8417 & 0.8464 & {\ul 0.8667} & 0.8312 & 0.7702 & 0.8374 & 0.7055 & 0.8485 & \textbf{0.8873} \\
&NDCG@2 & 0.6444 & 0.6101 & 0.6397 & 0.6373 & 0.6268 & 0.6311 & 0.6224 & {\ul 0.6754} & 0.6248 & 0.5010 & 0.6039 & 0.4258 & {\ul 0.6631} & \textbf{0.7143} \\
&NDCG@4 & 0.7159 & 0.6883 & 0.7061 & 0.7079 & 0.6967 & 0.7036 & 0.6983 & {\ul 0.7397} & 0.6953 & 0.5964 & 0.6845 & 0.5271 & {\ul 0.7263} & \textbf{0.7716} \\
&MRR & 0.6897 & 0.6623 & 0.6888 & 0.6851 & 0.6765 & 0.6818 & 0.6723 & {\ul 0.7177} & 0.6752 & 0.5736 & 0.6585 & 0.5121 & {\ul 0.7082} & \textbf{0.7501} \\
\bottomrule
\end{tabular}}\label{tab:pub_perf}
\end{table*}

\noindent\textbf{Industrial Dataset}.
For the industrial deployment on Kuaishou, we select these baseline methods performing well on the public Kuaishou dataset.
Specifically, the baseline methods and our method are deployed to the click-through rate (CTR) prediction module in the industrial recommendation engine.
The results are shown in Table \ref{tab:inds_perf}. Our FRec can improve both AUC and GAUC by more than 0.01 compared with the best baselines. This is a huge improvement for a real-world scenario with tens of millions of users~\cite{cheng2016wide,guo2017deepfm}. Note that this dataset has been scaled up hundreds of times, thus the improvement is more promising compared with that of the public Kuaishou dataset. We attribute this advantage to the choice of negative items in the evaluation. On the public Kuaishou dataset, negative items are randomly sampled. In contrast, we use the exposed videos users have not clicked on industrial dataset. In this scenario, user fatigue plays a critical role in consecutive decisions, where the exposed videos have generally matched user interests guaranteed by the advanced industrial recommender system. Therefore, FRec can effectively distinguish between clicked and non-clicked ones among all the exposed videos when modeling the influence of user fatigue on temporal interests accurately.

\begin{table}
\caption{Performance comparison on the industrial dataset.}
\begin{tabular}{c|c|c|c|c|c}
\toprule
Metric & GRU4Rec & SLi-Rec & CLSR & ComiRec-SA & FRec \\
\midrule
AUC & 0.7252 & 0.7302 & 0.7267 & 0.7247 & \textbf{0.7408} \\
GAUC & 0.6525 & 0.6604 & 0.6584 & 0.6433 & \textbf{0.6709} \\
\bottomrule
\end{tabular}
\label{tab:inds_perf}
\end{table}
\noindent\textbf{Efficiency Comparison}. Table \ref{tab:efficiency} shows the training time per epoch of all the models on public datasets, demonstrating that the efficiency of FRec is comparable with simple and complex baselines.

\begin{table*}[t]
\small
\centering
\caption{Training time~(minutes) per epoch of all the models.}
\begin{tabular}{c|cccccccc|cccc|cc}
\toprule
\multirow{2}{*}{Model} & \multirow{2}{*}{DIN} & \multirow{2}{*}{DIEN} & \multirow{2}{*}{GRU4Rec} & \multirow{2}{*}{SASRec} & \multirow{2}{*}{AdaMCT} & \multirow{2}{*}{Caser} & \multirow{2}{*}{SLi-Rec} & \multirow{2}{*}{CLSR} & \multirow{2}{*}{SUM} & \multirow{2}{*}{\shortstack{ComiRec\\-DR}} & \multirow{2}{*}{\shortstack{ComiRec\\-SA}} & \multirow{2}{*}{\shortstack{MGNM}}
 & \multirow{2}{*}{DFN} & \multirow{2}{*}{FRec} \\
  &&&&&&&&&&&&& \\ 
\midrule
\textbf{Kuaishou} & 17.0 & 17.2 & 18.8 & 59.3 & 17.8 & 16.8 & 24.1 & 21.7 & 83.2 & 16.6 & 17.0 & OOM & 19.5 & 23.2 \\
\textbf{Taobao} & 7.8 & 8.5 & 9.8 & 14.0 & 9.0 & 13.4 & 11.1 & 11.3 & 35.3 & 7.9 & 7.9 & 30.0 & 10.0 & 12.7 \\
\bottomrule
\end{tabular}\label{tab:efficiency}
\end{table*}

\subsection{Ablation Study~(RQ2)}
In our proposed FRec, there are some key modules for modeling user fatigue, including 1) fatigue-enhanced multi-interests fusion, 2) fatigue recurrent unit~(FRU) with fatigue representations as additional input, 3) cross networks for feature interplay to handle complex fatigue influence on user interests, and 4) contrastive learning with explicit fatigue supervision. To extensively verify their benefits, we conduct ablation studies to investigate how each module influences model effectiveness. Correspondingly, we have made the following changes,
\begin{itemize}[leftmargin=*]
    \item \textbf{w/o Fusion}: replace fatigue-enhanced attentive fusion with mean pooling.
    \item \textbf{w/o FRU}: replace FRU with vanilla GRU, with convolutional fatigue features removed.
    \item \textbf{w/o Cross}: replace cross layers with dense layers.
    \item \textbf{w/o CL}: remove contrastive learning, \textit{i.e.}, set $\alpha=0$.
\end{itemize}

\begin{figure*}[t!]
    \centering
    \subfigure[AUC on Kuaishou]{
    \includegraphics[scale=0.26]{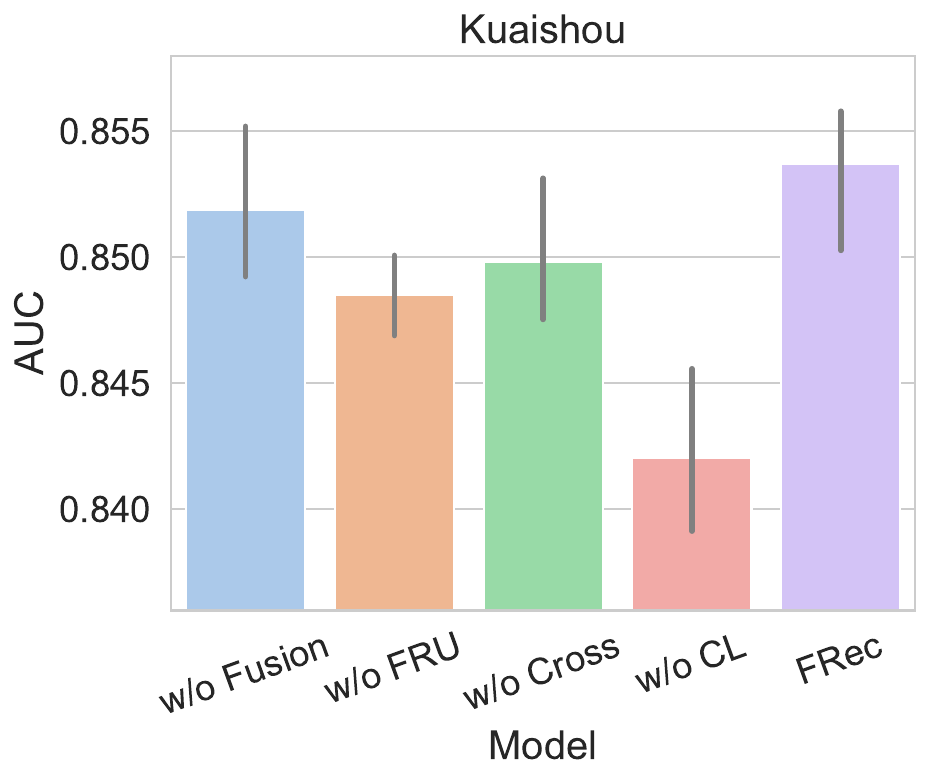}}
    \subfigure[NDCG@2 on Kuaishou]{
    \includegraphics[scale=0.26]{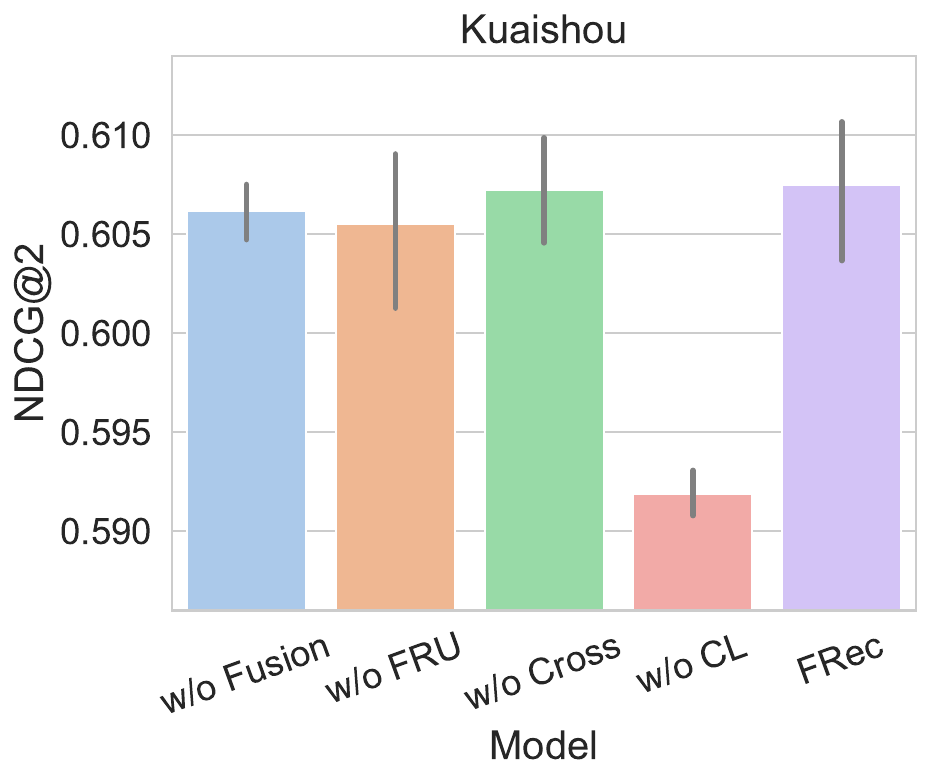}}
    \subfigure[AUC on Taobao]{
    \includegraphics[scale=0.26]{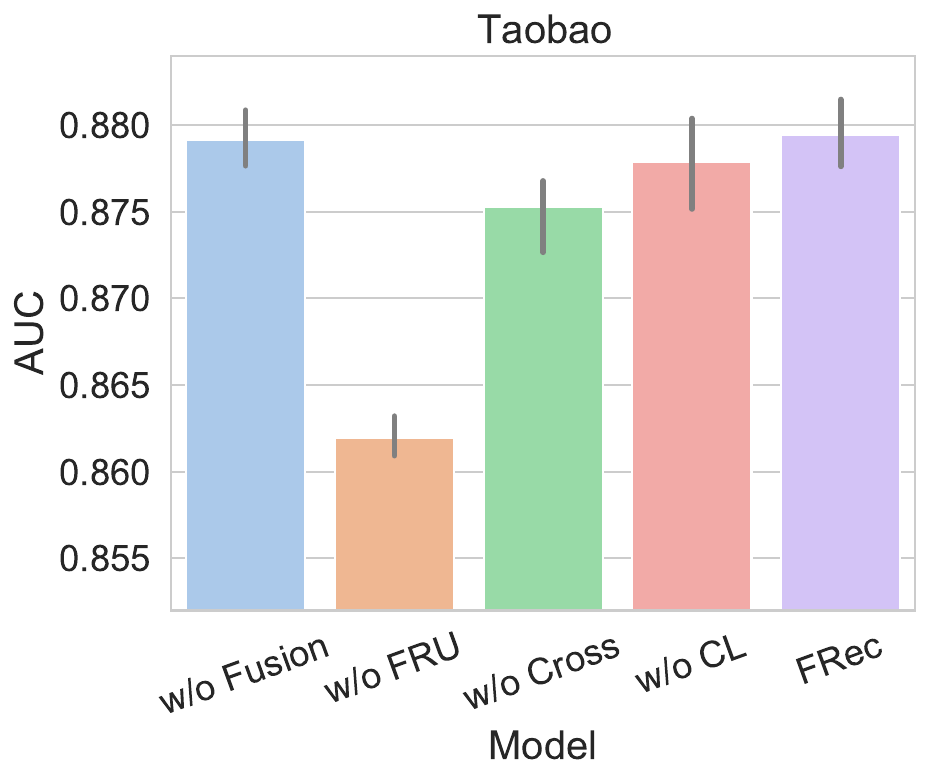}}
    \subfigure[NDCG@2 on Taobao]{
    \includegraphics[scale=0.26]{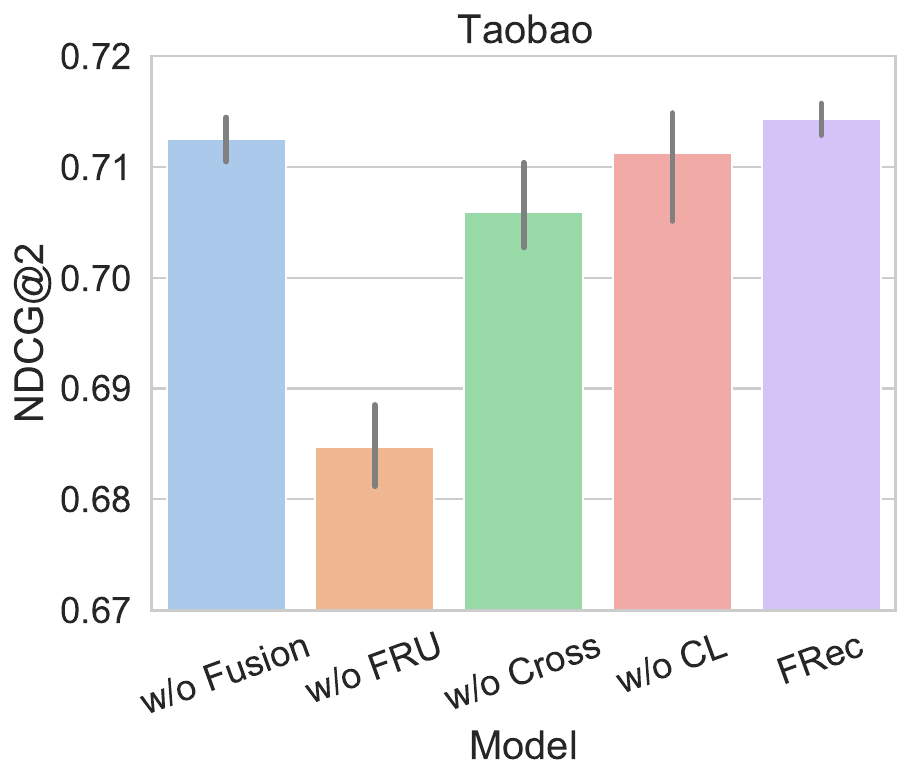}}
    \caption{Ablation study of key modules. Error bar denotes 95\% confidence intervals for five experiments.}
    \label{fig:ablation}
\end{figure*}

Performances with these key modules removed are shown in Figure \ref{fig:ablation}. Results show that FRec obtains better performances consistently compared with all the incomplete models on both datasets, demonstrating the necessity of the proposed modules for modeling user fatigue. Note that significantly worse performance without cross networks indicates critical benefits of the interplay of similarity features. 
Furthermore, on the Kuaishou dataset, the performance drops the most when contrastive learning is removed, but this is not the case for the Taobao dataset. We attribute this difference to the effectiveness of sequence augmentation for obtaining fatigue signals. In Kuaishou, a micro-video platform, repetitive recommendations of the same videos obviously cause intense user fatigue. However, for e-commerce platforms like Taobao, multiple views of the same product page are relatively common, and users may not experience fatigue during limited repetitions. In contrast, FRU plays an essential role in the model on the Taobao dataset. This demonstrates the effectiveness of guiding short-term interest evolution with temporal user fatigue as necessary inputs for update and reset gates. It can be explained by the intention changes of sequential behaviors in the e-commerce applications~\cite{chen2022intent,li2023multi}. Specifically, users have different intentions when browsing products, and they may experience fatigue of redundant exposure if the intention has switched. Therefore, fusing temporal user fatigue can assist in modeling short-term interests more accurately.

\subsection{Study on Fatigue Reduction~(RQ3)}

\noindent\textbf{Online Experiments}. We further deploy FRec on Kuaishou to verify the effectiveness of fatigue reduction and satisfaction improvement. Specifically, we choose CLSR~(the highest overall performance on public and industry datasets) for comparison and conduct an A/B test for 7 days, involving millions of users. Table \ref{tab:online} shows the improvement of key online metrics, which are defined as follows,
\begin{itemize}[leftmargin=*]
    \item \textbf{App usage} denotes the average dwell time users use the App.
    \item \textbf{\#Play} denotes total number of effectively played videos.
    \item \textbf{\#Category} denotes the average number of video categories in terms of the behavior of effective view.
    \item \textbf{Concentration} indicates how similar videos consecutively exposed over a fixed window are. It's calculated as $N-C$, where $N=6$ denotes the number of videos in the window, and $C$ is the number of video categories.
\end{itemize}
FRec improves all the metrics significantly around 0.1\%$\sim$0.4\%, which is impressive for large-scale online experiments. Obviously, FRec not only enables users to spend more time on the platform and view more videos but also promotes a more diverse video consumption. Particularly, the reduction in \textbf{Concentration} indicates that there are fewer videos of the same category in consecutive exposures of a short period, which lowers the perception of user fatigue.

\begin{figure}
    \centering
    \includegraphics[width=0.62\linewidth]{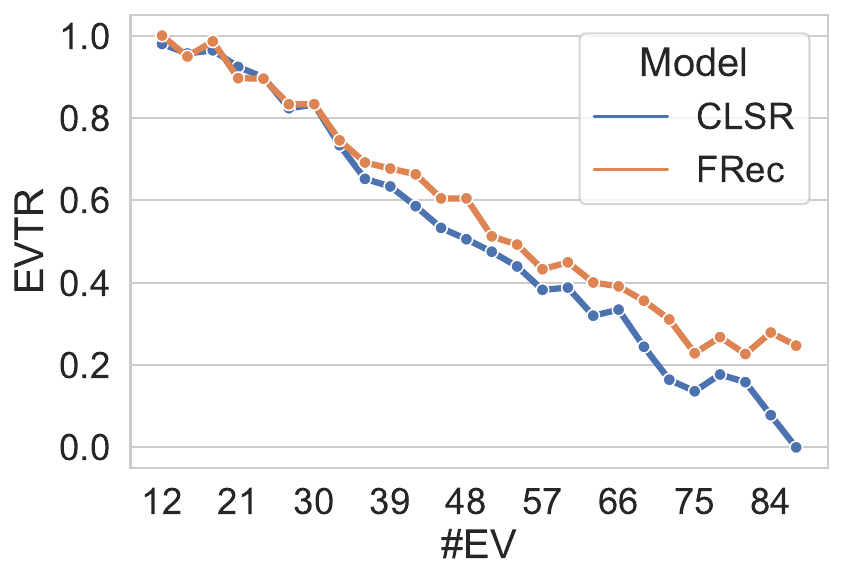}
    \caption{Online EVTR comparison between CLSR and FRec.}
    \label{fig:evtr_online}
\end{figure}

In order to further demonstrate the effectiveness of fatigue reduction, we conduct a similar analysis in Figure \ref{fig:evtr} with online interaction data. The comparison between the results of CLSR and FRec is shown in Figure \ref{fig:evtr_online}, where EVTR is also normalized. It's obvious that FRec can improve EVTR significantly when users have consumed many similar videos. Note that the improvement is obtained in an online setting, thus this is strong evidence of fatigue reduction. 

\begin{figure}[t]
    \centering
    \includegraphics[width=0.42\textwidth]{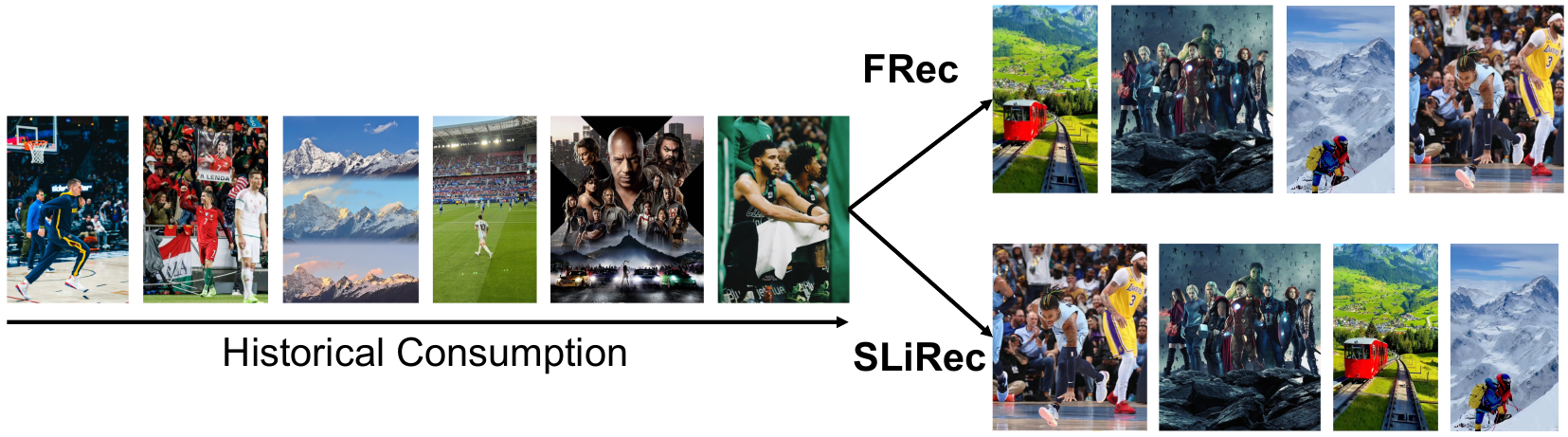}
    \caption{Recommendations by SLiRec and FRec.}
    \label{fig:case}
\end{figure}

\noindent\textbf{Offline Experiments}. Since it's not practical to conduct similar experiments on offline datasets, we conduct alternative experiments to demonstrate the fatigue reduction of FRec. We first show a recommendation case for a user in the industrial dataset. As illustrated in Figure \ref{fig:case}, the user has watched many sports videos recently, but SLi-Rec~(the best baseline) still ranks a basketball video in the first position. In contrast, our FRec ranks this video in the last position. This demonstrates that FRec can alleviate improper and repetitive recommendations that may cause user fatigue in a period of short time~(\textit{e.g.}, five minutes). Besides, as a part of user interests, the video about the scenery and trips lies at the top, with which the user has limited interactions. This demonstrates that FRec can assist in satisfying user interests and reducing user fatigue adaptively.

\begin{figure}[t]
    \centering
    \subfigure[AUC on Taobao]{
    \includegraphics[width=0.22\textwidth]{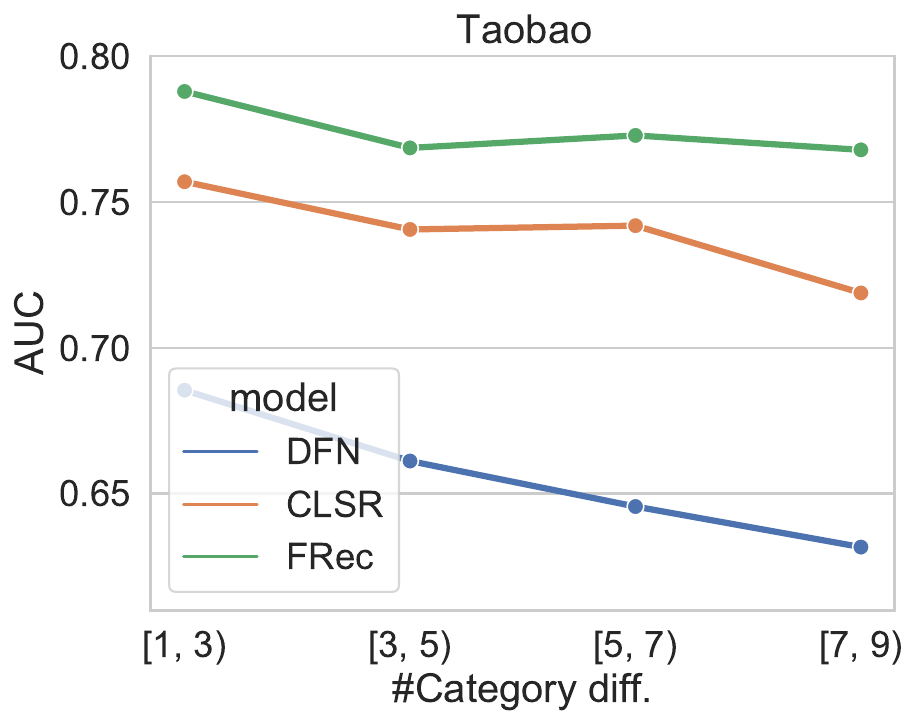}}
    \subfigure[NDCG@2 on Taobao]{
    \includegraphics[width=0.22\textwidth]{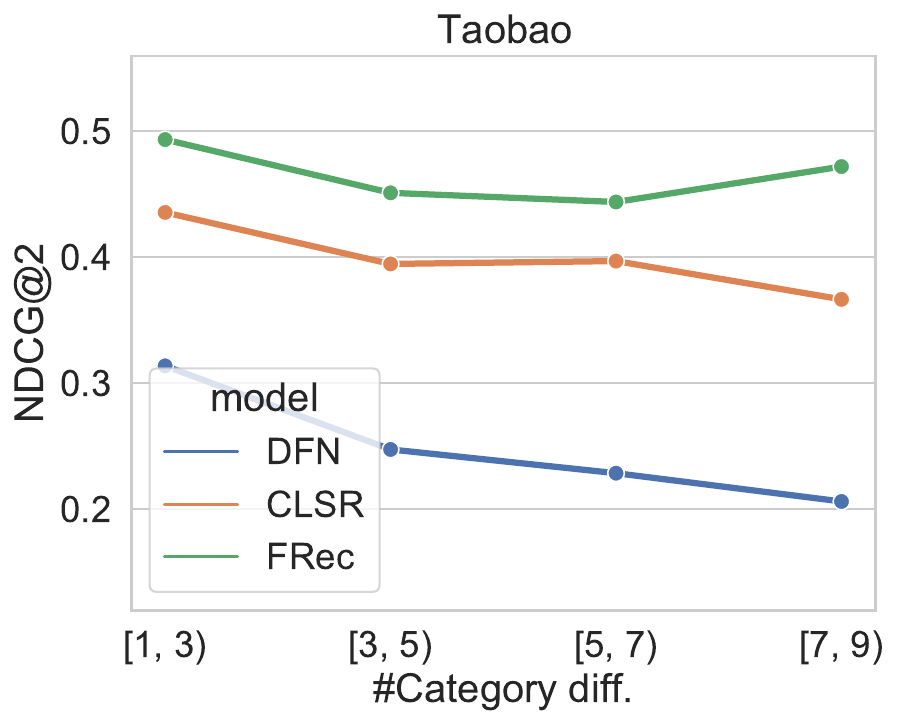}}
    \caption{Performance of instance groups with different $m$.}
    \label{fig:m}
\end{figure}

Furthermore, we investigate whether FRec can perform better when user fatigue plays an important role in user decisions. Specifically, we define a proxy to represent the importance, formulated as follows,
\begin{equation}
    m = \sum_{i_n} (m_{i_n}-m_{i_p}),
\end{equation}
where $m_{i_n}$~($m_{i_p}$) denotes the number of items within three-hour historical consumption that belong to the same category of the negative~(positive) item $i_n$~($i_p$). This proxy $m$ indicates the difference between historical-negative and historical-positive item similarity. High $m>0$ means that only modeling user interests based on relevance learning is insufficient for accurate recommendations. In other words, users are willing to interact with the item $i_p$ other than $i_n$ because of experiencing fatigue. We divide all the instances into groups with different $m$, and show the performance comparison in Figure \ref{fig:m}. Due to the sparsity of $m$ on the Kuaishou dataset, we only report results on the Taobao dataset. The best two baselines perform worse with increasing $m$, especially when $m\geq 5$. In contrast, our FRec keeps steady~(and significantly better) performances thanks to the ability to model users' temporal fatigue in short-term interest learning. Therefore, FRec can reduce user fatigue by ranking positive items at the top, which are less similar to historical items than negative items. 

\begin{table}
\caption{The improvement of key online metrics. $\uparrow$~($\downarrow$) means higher~(lower) is better.}
    \centering
    \small
    \begin{tabular}{c|c|c|c|c}
    \toprule
       \textbf{Metric}  & \textbf{App usage $\uparrow$} & \textbf{\#Play $\uparrow$} & \textbf{\#Category $\uparrow$} & \textbf{Concentration $\downarrow$}\\
       \hline
       Impr.~(\%)  & +0.300 & +0.466 & +0.408 & -0.136 \\
    \bottomrule
    \end{tabular}
    \label{tab:online}
\end{table}

\subsection{Hyper-parameter Study~(RQ4)}
We further investigate how key hyper-parameters impact the effectiveness of modeling user fatigue and recommendation performances. Due to the page limitations, we only report the results on the Taobao dataset. The results on the Kuaishou dataset imply the same conclusions.

\begin{figure}[t!]
    \centering
    \subfigure[Kernel size]{
    \includegraphics[width=0.23\textwidth]{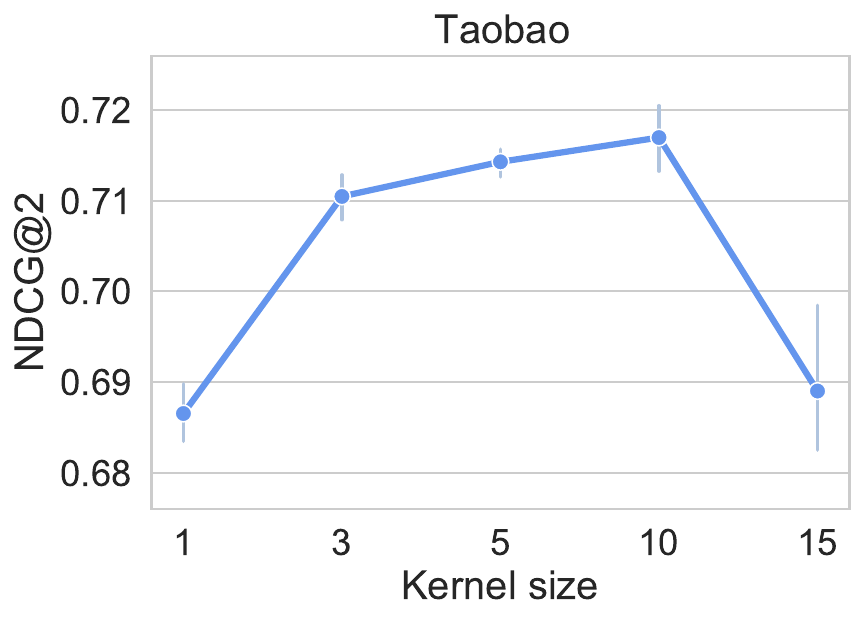}}
    \subfigure[Truncated threshold]{
    \includegraphics[width=0.23\textwidth]{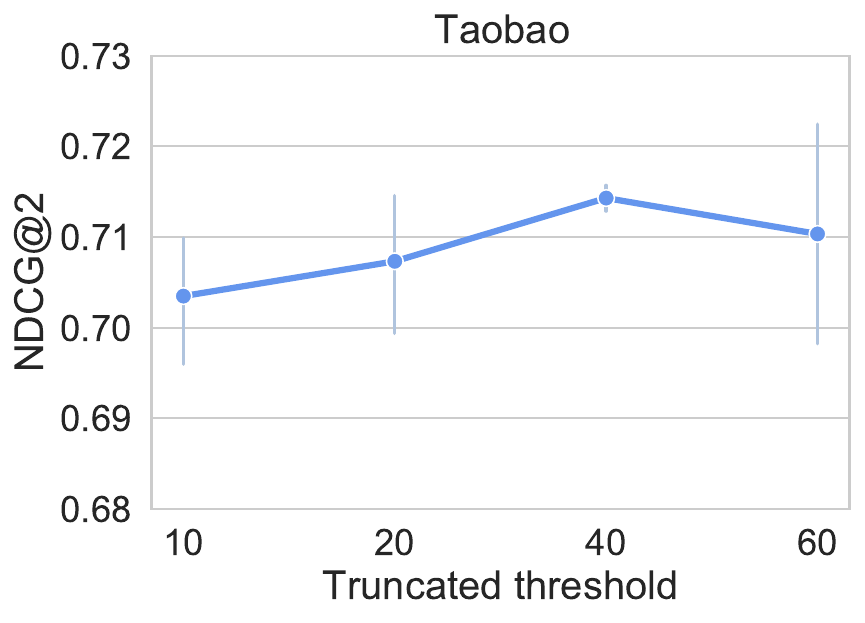}}
    \caption{Impact of kernel size and truncated threshold in 1D convolution. Error bar denotes 95\% confidence intervals for five experiments.}
    \label{fig:ksize}
\end{figure}

\textbf{Kernel size}. As stated in subsection \ref{fru}, considering consecutive items is necessary for modeling temporal fatigue. Therefore, we compare the performances under different kernel sizes of convolutions to verify whether FRec captures this pattern. As shown in Figure \ref{fig:ksize} (a), the performance is very low when kernel size is $1$ and is higher with a larger size. This is direct evidence that FRec can effectively model user fatigue caused by consecutive consumption. Note that the performance also drops when the size is too large, this may be explained by users' limited memory of historical experience.

\textbf{Truncated threshold}. From the results shown in Figure \ref{fig:ksize} (b), we observe that using too short~(10 items) or too long~(60 items) sub-sequence both leads to worse performances. This can be attributed to the missing of critical items and the inclusion of noisy items respectively, when modeling short-term interests and temporal user fatigue. Besides, from the comparison of confidence intervals, we conclude that the proper choice of the truncated threshold assists in obtaining stable recommendation results.

\section{Related Work}
\label{related}
\subsection{Sequential Recommendation}
In recent years, deep learning has been widely applied to sequential recommendations, and many advanced neural networks have been utilized for modeling long and short-term interests~\cite{li2017neural,sun2019bert4rec,an2019neural,li2020time}. Specifically, RNNs can directly capture the evolution of users' short-term interests when encoding sequential items, including gated recurrent unit~(GRU)~\cite{cho2014learning}, long short-term memory~(LSTM)~\cite{hochreiter1997long}, \textit{etc.} Similarly, CNNs have also been exploited to learn temporal patterns in historical consumption~\cite{tang2018personalized,xu2019recurrent,yan2019cosrec}. In terms of long-term interests, some works rely on matrix factorization or attention mechanisms~\cite{li2017neural,yu2019adaptive,li2020time}, which requires taking the whole sequence into consideration simultaneously. Recently, these two aspects have been disentangled for better modeling from the perspective of causal inference~\cite{zheng2022disentangling}.
Many works also propose to encode users' multiple interests by multiple representations simultaneously. In general, there are three types of modules for generating multiple interest representations, including multi-channel memory networks~\cite{pi2019practice,lian2021multi}, dynamic routing~\cite{li2019multi,cen2020controllable}, and self-attention mechanism~\cite{cen2020controllable}.

Different from these works, we take user fatigue into consideration and model its influence on long and short-term interests.

\subsection{User Fatigue in Recommendation}
Modeling user fatigue has not received much attention from the academic community of the recommender system. Several existing works~\cite{ma2016user,aharon2019soft,moriwaki2019fatigue,xie2022multi,li2023fan} rely on coarse-grained features representing how similar the target item is to historical consumption, such as 
how many historical items belong to the same category of the target item, \textit{etc.}
Then these features are directly feeded into the base recommender~(\textit{e.g.}, decision trees)~\cite{ma2016user,xie2022multi} or utilized for modeling fatigue by a quadratic function~\cite{moriwaki2019fatigue}. On one hand, these methods require manual feature engineering and enough features are difficult to obtain when relevant data is missing. On the other hand, the way to model user fatigue is not carefully designed to handle complex relationships with similarity features. 

In this work, we leverage fine-grained similarity features to support the modeling of user fatigue. Besides, user fatigue is also explicitly predicted based on contrastive learning.

\section{Conclusion and Future Work}
In this work, we propose to model user fatigue in interest learning for sequential recommendations. Specifically, based on a multi-interest framework, we develop an interest-aware similarity matrix for fatigue modeling and handle its influence on long and short-term user interests. We also propose a novel sequence augmentation to obtain fatigue signals as supervision for contrastive learning. Extensive offline and online experiments demonstrate the effectiveness of our model in improving user experience and reducing user fatigue. As for future work, we will propose introducing a fatigue metric as a new dimension to explicitly measure the effectiveness of recommendations, thereby encouraging the development of fatigue modeling in recommender system research.

\section*{Acknowledgement}
This work is supported by the National Natural Science Foundation of China under U23B2030, 62272262, and 72342032. This work is supported by a grant from the Guoqiang Institute, Tsinghua University under 2021GQG1005. This work is also supported by Kuaishou.

\clearpage
\bibliographystyle{ACM-Reference-Format}
\bibliography{sample-base}


\begin{thebibliography}{45}


\ifx \showCODEN    \undefined \def \showCODEN     #1{\unskip}     \fi
\ifx \showDOI      \undefined \def \showDOI       #1{#1}\fi
\ifx \showISBNx    \undefined \def \showISBNx     #1{\unskip}     \fi
\ifx \showISBNxiii \undefined \def \showISBNxiii  #1{\unskip}     \fi
\ifx \showISSN     \undefined \def \showISSN      #1{\unskip}     \fi
\ifx \showLCCN     \undefined \def \showLCCN      #1{\unskip}     \fi
\ifx \shownote     \undefined \def \shownote      #1{#1}          \fi
\ifx \showarticletitle \undefined \def \showarticletitle #1{#1}   \fi
\ifx \showURL      \undefined \def \showURL       {\relax}        \fi
\providecommand\bibfield[2]{#2}
\providecommand\bibinfo[2]{#2}
\providecommand\natexlab[1]{#1}
\providecommand\showeprint[2][]{arXiv:#2}

\bibitem[\protect\citeauthoryear{Aharon, Kaplan, Levy, Somekh, Blanc, Eshel, Shahar, Singer, and Zlotnik}{Aharon et~al\mbox{.}}{2019}]%
        {aharon2019soft}
\bibfield{author}{\bibinfo{person}{Michal Aharon}, \bibinfo{person}{Yohay Kaplan}, \bibinfo{person}{Rina Levy}, \bibinfo{person}{Oren Somekh}, \bibinfo{person}{Ayelet Blanc}, \bibinfo{person}{Neetai Eshel}, \bibinfo{person}{Avi Shahar}, \bibinfo{person}{Assaf Singer}, {and} \bibinfo{person}{Alex Zlotnik}.} \bibinfo{year}{2019}\natexlab{}.
\newblock \showarticletitle{Soft frequency capping for improved ad click prediction in yahoo gemini native}. In \bibinfo{booktitle}{{\em Proceedings of the 28th ACM International Conference on Information and Knowledge Management}}. \bibinfo{pages}{2793--2801}.
\newblock


\bibitem[\protect\citeauthoryear{Alhijawi, Awajan, and Fraihat}{Alhijawi et~al\mbox{.}}{2022}]%
        {alhijawi2022survey}
\bibfield{author}{\bibinfo{person}{Bushra Alhijawi}, \bibinfo{person}{Arafat Awajan}, {and} \bibinfo{person}{Salam Fraihat}.} \bibinfo{year}{2022}\natexlab{}.
\newblock \showarticletitle{Survey on the objectives of recommender systems: Measures, solutions, evaluation methodology, and new perspectives}.
\newblock \bibinfo{journal}{{\it Comput. Surveys}} \bibinfo{volume}{55}, \bibinfo{number}{5} (\bibinfo{year}{2022}), \bibinfo{pages}{1--38}.
\newblock


\bibitem[\protect\citeauthoryear{An, Wu, Wu, Zhang, Liu, and Xie}{An et~al\mbox{.}}{2019}]%
        {an2019neural}
\bibfield{author}{\bibinfo{person}{Mingxiao An}, \bibinfo{person}{Fangzhao Wu}, \bibinfo{person}{Chuhan Wu}, \bibinfo{person}{Kun Zhang}, \bibinfo{person}{Zheng Liu}, {and} \bibinfo{person}{Xing Xie}.} \bibinfo{year}{2019}\natexlab{}.
\newblock \showarticletitle{Neural news recommendation with long-and short-term user representations}. In \bibinfo{booktitle}{{\em Proceedings of the 57th Annual Meeting of the Association for Computational Linguistics}}. \bibinfo{pages}{336--345}.
\newblock


\bibitem[\protect\citeauthoryear{Bai, Kolter, and Koltun}{Bai et~al\mbox{.}}{2018}]%
        {bai2018empirical}
\bibfield{author}{\bibinfo{person}{Shaojie Bai}, \bibinfo{person}{J~Zico Kolter}, {and} \bibinfo{person}{Vladlen Koltun}.} \bibinfo{year}{2018}\natexlab{}.
\newblock \showarticletitle{An empirical evaluation of generic convolutional and recurrent networks for sequence modeling}.
\newblock \bibinfo{journal}{{\em arXiv preprint arXiv:1803.01271\/}} (\bibinfo{year}{2018}).
\newblock


\bibitem[\protect\citeauthoryear{Cen, Zhang, Zou, Zhou, Yang, and Tang}{Cen et~al\mbox{.}}{2020}]%
        {cen2020controllable}
\bibfield{author}{\bibinfo{person}{Yukuo Cen}, \bibinfo{person}{Jianwei Zhang}, \bibinfo{person}{Xu Zou}, \bibinfo{person}{Chang Zhou}, \bibinfo{person}{Hongxia Yang}, {and} \bibinfo{person}{Jie Tang}.} \bibinfo{year}{2020}\natexlab{}.
\newblock \showarticletitle{Controllable multi-interest framework for recommendation}. In \bibinfo{booktitle}{{\em Proceedings of the 26th ACM SIGKDD International Conference on Knowledge Discovery \& Data Mining}}. \bibinfo{pages}{2942--2951}.
\newblock


\bibitem[\protect\citeauthoryear{Chang, Gao, Zheng, Hui, Niu, Song, Jin, and Li}{Chang et~al\mbox{.}}{2021}]%
        {chang2021sequential}
\bibfield{author}{\bibinfo{person}{Jianxin Chang}, \bibinfo{person}{Chen Gao}, \bibinfo{person}{Yu Zheng}, \bibinfo{person}{Yiqun Hui}, \bibinfo{person}{Yanan Niu}, \bibinfo{person}{Yang Song}, \bibinfo{person}{Depeng Jin}, {and} \bibinfo{person}{Yong Li}.} \bibinfo{year}{2021}\natexlab{}.
\newblock \showarticletitle{Sequential Recommendation with Graph Neural Networks}. In \bibinfo{booktitle}{{\em Proceedings of the 44th International ACM SIGIR Conference on Research and Development in Information Retrieval}}. \bibinfo{pages}{378--387}.
\newblock


\bibitem[\protect\citeauthoryear{Chen, Liu, Li, McAuley, and Xiong}{Chen et~al\mbox{.}}{2022}]%
        {chen2022intent}
\bibfield{author}{\bibinfo{person}{Yongjun Chen}, \bibinfo{person}{Zhiwei Liu}, \bibinfo{person}{Jia Li}, \bibinfo{person}{Julian McAuley}, {and} \bibinfo{person}{Caiming Xiong}.} \bibinfo{year}{2022}\natexlab{}.
\newblock \showarticletitle{Intent contrastive learning for sequential recommendation}. In \bibinfo{booktitle}{{\em Proceedings of the ACM Web Conference 2022}}. \bibinfo{pages}{2172--2182}.
\newblock


\bibitem[\protect\citeauthoryear{Cheng, Koc, Harmsen, Shaked, Chandra, Aradhye, Anderson, Corrado, Chai, Ispir, et~al\mbox{.}}{Cheng et~al\mbox{.}}{2016}]%
        {cheng2016wide}
\bibfield{author}{\bibinfo{person}{Heng-Tze Cheng}, \bibinfo{person}{Levent Koc}, \bibinfo{person}{Jeremiah Harmsen}, \bibinfo{person}{Tal Shaked}, \bibinfo{person}{Tushar Chandra}, \bibinfo{person}{Hrishi Aradhye}, \bibinfo{person}{Glen Anderson}, \bibinfo{person}{Greg Corrado}, \bibinfo{person}{Wei Chai}, \bibinfo{person}{Mustafa Ispir}, {et~al\mbox{.}}} \bibinfo{year}{2016}\natexlab{}.
\newblock \showarticletitle{Wide \& deep learning for recommender systems}. In \bibinfo{booktitle}{{\em Proceedings of the 1st workshop on deep learning for recommender systems}}. \bibinfo{pages}{7--10}.
\newblock


\bibitem[\protect\citeauthoryear{Cho, Van~Merri{\"e}nboer, Gulcehre, Bahdanau, Bougares, Schwenk, and Bengio}{Cho et~al\mbox{.}}{2014}]%
        {cho2014learning}
\bibfield{author}{\bibinfo{person}{Kyunghyun Cho}, \bibinfo{person}{Bart Van~Merri{\"e}nboer}, \bibinfo{person}{Caglar Gulcehre}, \bibinfo{person}{Dzmitry Bahdanau}, \bibinfo{person}{Fethi Bougares}, \bibinfo{person}{Holger Schwenk}, {and} \bibinfo{person}{Yoshua Bengio}.} \bibinfo{year}{2014}\natexlab{}.
\newblock \showarticletitle{Learning phrase representations using RNN encoder-decoder for statistical machine translation}.
\newblock \bibinfo{journal}{{\em arXiv preprint arXiv:1406.1078\/}} (\bibinfo{year}{2014}).
\newblock


\bibitem[\protect\citeauthoryear{Ding, Quan, He, Li, and Jin}{Ding et~al\mbox{.}}{2019}]%
        {ding2019reinforced}
\bibfield{author}{\bibinfo{person}{Jingtao Ding}, \bibinfo{person}{Yuhan Quan}, \bibinfo{person}{Xiangnan He}, \bibinfo{person}{Yong Li}, {and} \bibinfo{person}{Depeng Jin}.} \bibinfo{year}{2019}\natexlab{}.
\newblock \showarticletitle{Reinforced Negative Sampling for Recommendation with Exposure Data.}. In \bibinfo{booktitle}{{\em IJCAI}}. Macao, \bibinfo{pages}{2230--2236}.
\newblock


\bibitem[\protect\citeauthoryear{Ding, Quan, Yao, Li, and Jin}{Ding et~al\mbox{.}}{2020}]%
        {ding2020simplify}
\bibfield{author}{\bibinfo{person}{Jingtao Ding}, \bibinfo{person}{Yuhan Quan}, \bibinfo{person}{Quanming Yao}, \bibinfo{person}{Yong Li}, {and} \bibinfo{person}{Depeng Jin}.} \bibinfo{year}{2020}\natexlab{}.
\newblock \showarticletitle{Simplify and robustify negative sampling for implicit collaborative filtering}.
\newblock \bibinfo{journal}{{\em Advances in Neural Information Processing Systems\/}}  \bibinfo{volume}{33} (\bibinfo{year}{2020}), \bibinfo{pages}{1094--1105}.
\newblock


\bibitem[\protect\citeauthoryear{Fu, Niu, and Maher}{Fu et~al\mbox{.}}{2023}]%
        {fu2023deep}
\bibfield{author}{\bibinfo{person}{Zhe Fu}, \bibinfo{person}{Xi Niu}, {and} \bibinfo{person}{Mary~Lou Maher}.} \bibinfo{year}{2023}\natexlab{}.
\newblock \showarticletitle{Deep learning models for serendipity recommendations: a survey and new perspectives}.
\newblock \bibinfo{journal}{{\it Comput. Surveys}} \bibinfo{volume}{56}, \bibinfo{number}{1} (\bibinfo{year}{2023}), \bibinfo{pages}{1--26}.
\newblock


\bibitem[\protect\citeauthoryear{Gao, Zheng, Wang, Feng, He, and Li}{Gao et~al\mbox{.}}{2024}]%
        {gao2024causal}
\bibfield{author}{\bibinfo{person}{Chen Gao}, \bibinfo{person}{Yu Zheng}, \bibinfo{person}{Wenjie Wang}, \bibinfo{person}{Fuli Feng}, \bibinfo{person}{Xiangnan He}, {and} \bibinfo{person}{Yong Li}.} \bibinfo{year}{2024}\natexlab{}.
\newblock \showarticletitle{Causal inference in recommender systems: A survey and future directions}.
\newblock \bibinfo{journal}{{\em ACM Transactions on Information Systems\/}} \bibinfo{volume}{42}, \bibinfo{number}{4} (\bibinfo{year}{2024}), \bibinfo{pages}{1--32}.
\newblock


\bibitem[\protect\citeauthoryear{Guo, Tang, Ye, Li, and He}{Guo et~al\mbox{.}}{2017}]%
        {guo2017deepfm}
\bibfield{author}{\bibinfo{person}{Huifeng Guo}, \bibinfo{person}{Ruiming Tang}, \bibinfo{person}{Yunming Ye}, \bibinfo{person}{Zhenguo Li}, {and} \bibinfo{person}{Xiuqiang He}.} \bibinfo{year}{2017}\natexlab{}.
\newblock \showarticletitle{DeepFM: a factorization-machine based neural network for CTR prediction}.
\newblock \bibinfo{journal}{{\em arXiv preprint arXiv:1703.04247\/}} (\bibinfo{year}{2017}).
\newblock


\bibitem[\protect\citeauthoryear{Hamilton, Ying, and Leskovec}{Hamilton et~al\mbox{.}}{2017}]%
        {hamilton2017inductive}
\bibfield{author}{\bibinfo{person}{Will Hamilton}, \bibinfo{person}{Zhitao Ying}, {and} \bibinfo{person}{Jure Leskovec}.} \bibinfo{year}{2017}\natexlab{}.
\newblock \showarticletitle{Inductive representation learning on large graphs}.
\newblock \bibinfo{journal}{{\em Advances in neural information processing systems\/}}  \bibinfo{volume}{30} (\bibinfo{year}{2017}).
\newblock


\bibitem[\protect\citeauthoryear{Hidasi, Karatzoglou, Baltrunas, and Tikk}{Hidasi et~al\mbox{.}}{2015}]%
        {hidasi2015session}
\bibfield{author}{\bibinfo{person}{Bal{\'a}zs Hidasi}, \bibinfo{person}{Alexandros Karatzoglou}, \bibinfo{person}{Linas Baltrunas}, {and} \bibinfo{person}{Domonkos Tikk}.} \bibinfo{year}{2015}\natexlab{}.
\newblock \showarticletitle{Session-based recommendations with recurrent neural networks}.
\newblock \bibinfo{journal}{{\em arXiv preprint arXiv:1511.06939\/}} (\bibinfo{year}{2015}).
\newblock


\bibitem[\protect\citeauthoryear{Hochreiter and Schmidhuber}{Hochreiter and Schmidhuber}{1997}]%
        {hochreiter1997long}
\bibfield{author}{\bibinfo{person}{Sepp Hochreiter} {and} \bibinfo{person}{J{\"u}rgen Schmidhuber}.} \bibinfo{year}{1997}\natexlab{}.
\newblock \showarticletitle{Long short-term memory}.
\newblock \bibinfo{journal}{{\em Neural computation\/}} \bibinfo{volume}{9}, \bibinfo{number}{8} (\bibinfo{year}{1997}), \bibinfo{pages}{1735--1780}.
\newblock


\bibitem[\protect\citeauthoryear{Jiang, Zhang, Luo, Li, Kim, Zhang, Wang, Xie, and Kim}{Jiang et~al\mbox{.}}{2023}]%
        {jiang2023adamct}
\bibfield{author}{\bibinfo{person}{Juyong Jiang}, \bibinfo{person}{Peiyan Zhang}, \bibinfo{person}{Yingtao Luo}, \bibinfo{person}{Chaozhuo Li}, \bibinfo{person}{Jae~Boum Kim}, \bibinfo{person}{Kai Zhang}, \bibinfo{person}{Senzhang Wang}, \bibinfo{person}{Xing Xie}, {and} \bibinfo{person}{Sunghun Kim}.} \bibinfo{year}{2023}\natexlab{}.
\newblock \showarticletitle{AdaMCT: adaptive mixture of CNN-transformer for sequential recommendation}. In \bibinfo{booktitle}{{\em Proceedings of the 32nd ACM International Conference on Information and Knowledge Management}}. \bibinfo{pages}{976--986}.
\newblock


\bibitem[\protect\citeauthoryear{Kang and McAuley}{Kang and McAuley}{2018}]%
        {kang2018self}
\bibfield{author}{\bibinfo{person}{Wang-Cheng Kang} {and} \bibinfo{person}{Julian McAuley}.} \bibinfo{year}{2018}\natexlab{}.
\newblock \showarticletitle{Self-attentive sequential recommendation}. In \bibinfo{booktitle}{{\em 2018 IEEE International Conference on Data Mining (ICDM)}}. IEEE, \bibinfo{pages}{197--206}.
\newblock


\bibitem[\protect\citeauthoryear{Kingma and Ba}{Kingma and Ba}{2014}]%
        {kingma2014adam}
\bibfield{author}{\bibinfo{person}{Diederik~P Kingma} {and} \bibinfo{person}{Jimmy Ba}.} \bibinfo{year}{2014}\natexlab{}.
\newblock \showarticletitle{Adam: A method for stochastic optimization}.
\newblock \bibinfo{journal}{{\em arXiv preprint arXiv:1412.6980\/}} (\bibinfo{year}{2014}).
\newblock


\bibitem[\protect\citeauthoryear{Li, Liu, Wu, Xu, Zhao, Huang, Kang, Chen, Li, and Lee}{Li et~al\mbox{.}}{2019}]%
        {li2019multi}
\bibfield{author}{\bibinfo{person}{Chao Li}, \bibinfo{person}{Zhiyuan Liu}, \bibinfo{person}{Mengmeng Wu}, \bibinfo{person}{Yuchi Xu}, \bibinfo{person}{Huan Zhao}, \bibinfo{person}{Pipei Huang}, \bibinfo{person}{Guoliang Kang}, \bibinfo{person}{Qiwei Chen}, \bibinfo{person}{Wei Li}, {and} \bibinfo{person}{Dik~Lun Lee}.} \bibinfo{year}{2019}\natexlab{}.
\newblock \showarticletitle{Multi-interest network with dynamic routing for recommendation at Tmall}. In \bibinfo{booktitle}{{\em Proceedings of the 28th ACM international conference on information and knowledge management}}. \bibinfo{pages}{2615--2623}.
\newblock


\bibitem[\protect\citeauthoryear{Li, Ren, Chen, Ren, Lian, and Ma}{Li et~al\mbox{.}}{2017}]%
        {li2017neural}
\bibfield{author}{\bibinfo{person}{Jing Li}, \bibinfo{person}{Pengjie Ren}, \bibinfo{person}{Zhumin Chen}, \bibinfo{person}{Zhaochun Ren}, \bibinfo{person}{Tao Lian}, {and} \bibinfo{person}{Jun Ma}.} \bibinfo{year}{2017}\natexlab{}.
\newblock \showarticletitle{Neural attentive session-based recommendation}. In \bibinfo{booktitle}{{\em Proceedings of the 2017 ACM on Conference on Information and Knowledge Management}}. \bibinfo{pages}{1419--1428}.
\newblock


\bibitem[\protect\citeauthoryear{Li, Wang, and McAuley}{Li et~al\mbox{.}}{2020}]%
        {li2020time}
\bibfield{author}{\bibinfo{person}{Jiacheng Li}, \bibinfo{person}{Yujie Wang}, {and} \bibinfo{person}{Julian McAuley}.} \bibinfo{year}{2020}\natexlab{}.
\newblock \showarticletitle{Time interval aware self-attention for sequential recommendation}. In \bibinfo{booktitle}{{\em Proceedings of the 13th international conference on web search and data mining}}. \bibinfo{pages}{322--330}.
\newblock


\bibitem[\protect\citeauthoryear{Li, Liu, Pan, Huang, Li, Su, Mao, and Cao}{Li et~al\mbox{.}}{2023a}]%
        {li2023fan}
\bibfield{author}{\bibinfo{person}{Ming Li}, \bibinfo{person}{Naiyin Liu}, \bibinfo{person}{Xiaofeng Pan}, \bibinfo{person}{Yang Huang}, \bibinfo{person}{Ningning Li}, \bibinfo{person}{Yingmin Su}, \bibinfo{person}{Chengjun Mao}, {and} \bibinfo{person}{Bo Cao}.} \bibinfo{year}{2023}\natexlab{a}.
\newblock \showarticletitle{FAN: Fatigue-Aware Network for Click-Through Rate Prediction in E-commerce Recommendation}. In \bibinfo{booktitle}{{\em Database Systems for Advanced Applications: 28th International Conference, DASFAA 2023, Tianjin, China, April 17--20, 2023, Proceedings, Part IV}}. Springer, \bibinfo{pages}{502--514}.
\newblock


\bibitem[\protect\citeauthoryear{Li, Sun, Zhao, Yu, Zhu, Jin, Yu, and Yu}{Li et~al\mbox{.}}{2023b}]%
        {li2023multi}
\bibfield{author}{\bibinfo{person}{Xuewei Li}, \bibinfo{person}{Aitong Sun}, \bibinfo{person}{Mankun Zhao}, \bibinfo{person}{Jian Yu}, \bibinfo{person}{Kun Zhu}, \bibinfo{person}{Di Jin}, \bibinfo{person}{Mei Yu}, {and} \bibinfo{person}{Ruiguo Yu}.} \bibinfo{year}{2023}\natexlab{b}.
\newblock \showarticletitle{Multi-Intention Oriented Contrastive Learning for Sequential Recommendation}. In \bibinfo{booktitle}{{\em Proceedings of the Sixteenth ACM International Conference on Web Search and Data Mining}}. \bibinfo{pages}{411--419}.
\newblock


\bibitem[\protect\citeauthoryear{Lian, Batal, Liu, Soni, Kang, Wang, and Xie}{Lian et~al\mbox{.}}{2021}]%
        {lian2021multi}
\bibfield{author}{\bibinfo{person}{Jianxun Lian}, \bibinfo{person}{Iyad Batal}, \bibinfo{person}{Zheng Liu}, \bibinfo{person}{Akshay Soni}, \bibinfo{person}{Eun~Yong Kang}, \bibinfo{person}{Yajun Wang}, {and} \bibinfo{person}{Xing Xie}.} \bibinfo{year}{2021}\natexlab{}.
\newblock \showarticletitle{Multi-Interest-Aware User Modeling for Large-Scale Sequential Recommendations}.
\newblock \bibinfo{journal}{{\em arXiv preprint arXiv:2102.09211\/}} (\bibinfo{year}{2021}).
\newblock


\bibitem[\protect\citeauthoryear{Lin, Gao, Li, Zheng, Li, Jin, and Li}{Lin et~al\mbox{.}}{2022}]%
        {lin2022dual}
\bibfield{author}{\bibinfo{person}{Guanyu Lin}, \bibinfo{person}{Chen Gao}, \bibinfo{person}{Yinfeng Li}, \bibinfo{person}{Yu Zheng}, \bibinfo{person}{Zhiheng Li}, \bibinfo{person}{Depeng Jin}, {and} \bibinfo{person}{Yong Li}.} \bibinfo{year}{2022}\natexlab{}.
\newblock \showarticletitle{Dual contrastive network for sequential recommendation}. In \bibinfo{booktitle}{{\em Proceedings of the 45th international ACM SIGIR conference on research and development in information retrieval}}. \bibinfo{pages}{2686--2691}.
\newblock


\bibitem[\protect\citeauthoryear{Lin, Feng, Santos, Yu, Xiang, Zhou, and Bengio}{Lin et~al\mbox{.}}{2017}]%
        {lin2017structured}
\bibfield{author}{\bibinfo{person}{Zhouhan Lin}, \bibinfo{person}{Minwei Feng}, \bibinfo{person}{Cicero Nogueira~dos Santos}, \bibinfo{person}{Mo Yu}, \bibinfo{person}{Bing Xiang}, \bibinfo{person}{Bowen Zhou}, {and} \bibinfo{person}{Yoshua Bengio}.} \bibinfo{year}{2017}\natexlab{}.
\newblock \showarticletitle{A structured self-attentive sentence embedding}.
\newblock \bibinfo{journal}{{\em arXiv preprint arXiv:1703.03130\/}} (\bibinfo{year}{2017}).
\newblock


\bibitem[\protect\citeauthoryear{Ma, Liu, and Shen}{Ma et~al\mbox{.}}{2016}]%
        {ma2016user}
\bibfield{author}{\bibinfo{person}{Hao Ma}, \bibinfo{person}{Xueqing Liu}, {and} \bibinfo{person}{Zhihong Shen}.} \bibinfo{year}{2016}\natexlab{}.
\newblock \showarticletitle{User fatigue in online news recommendation}. In \bibinfo{booktitle}{{\em Proceedings of the 25th International Conference on World Wide Web}}. \bibinfo{pages}{1363--1372}.
\newblock


\bibitem[\protect\citeauthoryear{Moriwaki, Fujita, Yasui, and Hoshino}{Moriwaki et~al\mbox{.}}{2019}]%
        {moriwaki2019fatigue}
\bibfield{author}{\bibinfo{person}{Daisuke Moriwaki}, \bibinfo{person}{Komei Fujita}, \bibinfo{person}{Shota Yasui}, {and} \bibinfo{person}{Takahiro Hoshino}.} \bibinfo{year}{2019}\natexlab{}.
\newblock \showarticletitle{Fatigue-Aware Ad Creative Selection}.
\newblock \bibinfo{journal}{{\em arXiv preprint arXiv:1908.08936\/}} (\bibinfo{year}{2019}).
\newblock


\bibitem[\protect\citeauthoryear{Pi, Bian, Zhou, Zhu, and Gai}{Pi et~al\mbox{.}}{2019}]%
        {pi2019practice}
\bibfield{author}{\bibinfo{person}{Qi Pi}, \bibinfo{person}{Weijie Bian}, \bibinfo{person}{Guorui Zhou}, \bibinfo{person}{Xiaoqiang Zhu}, {and} \bibinfo{person}{Kun Gai}.} \bibinfo{year}{2019}\natexlab{}.
\newblock \showarticletitle{Practice on long sequential user behavior modeling for click-through rate prediction}. In \bibinfo{booktitle}{{\em Proceedings of the 25th ACM SIGKDD International Conference on Knowledge Discovery \& Data Mining}}. \bibinfo{pages}{2671--2679}.
\newblock


\bibitem[\protect\citeauthoryear{Quan, Ding, Gao, Li, Yi, Jin, and Li}{Quan et~al\mbox{.}}{2023a}]%
        {quan2023alleviating}
\bibfield{author}{\bibinfo{person}{Yuhan Quan}, \bibinfo{person}{Jingtao Ding}, \bibinfo{person}{Chen Gao}, \bibinfo{person}{Nian Li}, \bibinfo{person}{Lingling Yi}, \bibinfo{person}{Depeng Jin}, {and} \bibinfo{person}{Yong Li}.} \bibinfo{year}{2023}\natexlab{a}.
\newblock \showarticletitle{Alleviating Video-length Effect for Micro-video Recommendation}.
\newblock \bibinfo{journal}{{\em ACM Transactions on Information Systems\/}} \bibinfo{volume}{42}, \bibinfo{number}{2} (\bibinfo{year}{2023}), \bibinfo{pages}{1--24}.
\newblock


\bibitem[\protect\citeauthoryear{Quan, Ding, Gao, Yi, Jin, and Li}{Quan et~al\mbox{.}}{2023b}]%
        {quan2023robust}
\bibfield{author}{\bibinfo{person}{Yuhan Quan}, \bibinfo{person}{Jingtao Ding}, \bibinfo{person}{Chen Gao}, \bibinfo{person}{Lingling Yi}, \bibinfo{person}{Depeng Jin}, {and} \bibinfo{person}{Yong Li}.} \bibinfo{year}{2023}\natexlab{b}.
\newblock \showarticletitle{Robust preference-guided denoising for graph based social recommendation}. In \bibinfo{booktitle}{{\em Proceedings of the ACM Web Conference 2023}}. \bibinfo{pages}{1097--1108}.
\newblock


\bibitem[\protect\citeauthoryear{Sun, Liu, Wu, Pei, Lin, Ou, and Jiang}{Sun et~al\mbox{.}}{2019}]%
        {sun2019bert4rec}
\bibfield{author}{\bibinfo{person}{Fei Sun}, \bibinfo{person}{Jun Liu}, \bibinfo{person}{Jian Wu}, \bibinfo{person}{Changhua Pei}, \bibinfo{person}{Xiao Lin}, \bibinfo{person}{Wenwu Ou}, {and} \bibinfo{person}{Peng Jiang}.} \bibinfo{year}{2019}\natexlab{}.
\newblock \showarticletitle{BERT4Rec: Sequential recommendation with bidirectional encoder representations from transformer}. In \bibinfo{booktitle}{{\em Proceedings of the 28th ACM International Conference on Information and Knowledge Management}}. \bibinfo{pages}{1441--1450}.
\newblock


\bibitem[\protect\citeauthoryear{Tang and Wang}{Tang and Wang}{2018}]%
        {tang2018personalized}
\bibfield{author}{\bibinfo{person}{Jiaxi Tang} {and} \bibinfo{person}{Ke Wang}.} \bibinfo{year}{2018}\natexlab{}.
\newblock \showarticletitle{Personalized top-n sequential recommendation via convolutional sequence embedding}. In \bibinfo{booktitle}{{\em Proceedings of the Eleventh ACM International Conference on Web Search and Data Mining}}. \bibinfo{pages}{565--573}.
\newblock


\bibitem[\protect\citeauthoryear{Tian, Chang, Niu, Song, and Li}{Tian et~al\mbox{.}}{2022}]%
        {tian2022multi}
\bibfield{author}{\bibinfo{person}{Yu Tian}, \bibinfo{person}{Jianxin Chang}, \bibinfo{person}{Yanan Niu}, \bibinfo{person}{Yang Song}, {and} \bibinfo{person}{Chenliang Li}.} \bibinfo{year}{2022}\natexlab{}.
\newblock \showarticletitle{When multi-level meets multi-interest: A multi-grained neural model for sequential recommendation}. In \bibinfo{booktitle}{{\em Proceedings of the 45th International ACM SIGIR Conference on Research and Development in Information Retrieval}}. \bibinfo{pages}{1632--1641}.
\newblock


\bibitem[\protect\citeauthoryear{Wang, Fu, Fu, and Wang}{Wang et~al\mbox{.}}{2017}]%
        {wang2017deep}
\bibfield{author}{\bibinfo{person}{Ruoxi Wang}, \bibinfo{person}{Bin Fu}, \bibinfo{person}{Gang Fu}, {and} \bibinfo{person}{Mingliang Wang}.} \bibinfo{year}{2017}\natexlab{}.
\newblock \showarticletitle{Deep \& cross network for ad click predictions}.
\newblock In \bibinfo{booktitle}{{\em Proceedings of the ADKDD'17}}. \bibinfo{pages}{1--7}.
\newblock


\bibitem[\protect\citeauthoryear{Xie, Ling, Zhang, Xia, and Lin}{Xie et~al\mbox{.}}{2022}]%
        {xie2022multi}
\bibfield{author}{\bibinfo{person}{Ruobing Xie}, \bibinfo{person}{Cheng Ling}, \bibinfo{person}{Shaoliang Zhang}, \bibinfo{person}{Feng Xia}, {and} \bibinfo{person}{Leyu Lin}.} \bibinfo{year}{2022}\natexlab{}.
\newblock \showarticletitle{Multi-granularity Fatigue in Recommendation}. In \bibinfo{booktitle}{{\em Proceedings of the 31st ACM International Conference on Information \& Knowledge Management}}. \bibinfo{pages}{4595--4599}.
\newblock


\bibitem[\protect\citeauthoryear{Xu, Zhao, Liu, Xu, S.~Sheng, Cui, Zhou, and Xiong}{Xu et~al\mbox{.}}{2019}]%
        {xu2019recurrent}
\bibfield{author}{\bibinfo{person}{Chengfeng Xu}, \bibinfo{person}{Pengpeng Zhao}, \bibinfo{person}{Yanchi Liu}, \bibinfo{person}{Jiajie Xu}, \bibinfo{person}{Victor S~Sheng S.~Sheng}, \bibinfo{person}{Zhiming Cui}, \bibinfo{person}{Xiaofang Zhou}, {and} \bibinfo{person}{Hui Xiong}.} \bibinfo{year}{2019}\natexlab{}.
\newblock \showarticletitle{Recurrent convolutional neural network for sequential recommendation}. In \bibinfo{booktitle}{{\em The world wide web conference}}. \bibinfo{pages}{3398--3404}.
\newblock


\bibitem[\protect\citeauthoryear{Yan, Cheng, Kang, Wan, and McAuley}{Yan et~al\mbox{.}}{2019}]%
        {yan2019cosrec}
\bibfield{author}{\bibinfo{person}{An Yan}, \bibinfo{person}{Shuo Cheng}, \bibinfo{person}{Wang-Cheng Kang}, \bibinfo{person}{Mengting Wan}, {and} \bibinfo{person}{Julian McAuley}.} \bibinfo{year}{2019}\natexlab{}.
\newblock \showarticletitle{CosRec: 2D convolutional neural networks for sequential recommendation}. In \bibinfo{booktitle}{{\em Proceedings of the 28th ACM international conference on information and knowledge management}}. \bibinfo{pages}{2173--2176}.
\newblock


\bibitem[\protect\citeauthoryear{Yu, Yin, Xia, Chen, Li, and Huang}{Yu et~al\mbox{.}}{2022}]%
        {yu2022self}
\bibfield{author}{\bibinfo{person}{Junliang Yu}, \bibinfo{person}{Hongzhi Yin}, \bibinfo{person}{Xin Xia}, \bibinfo{person}{Tong Chen}, \bibinfo{person}{Jundong Li}, {and} \bibinfo{person}{Zi Huang}.} \bibinfo{year}{2022}\natexlab{}.
\newblock \showarticletitle{Self-supervised learning for recommender systems: A survey}.
\newblock \bibinfo{journal}{{\em arXiv preprint arXiv:2203.15876\/}} (\bibinfo{year}{2022}).
\newblock


\bibitem[\protect\citeauthoryear{Yu, Lian, Mahmoody, Liu, and Xie}{Yu et~al\mbox{.}}{2019}]%
        {yu2019adaptive}
\bibfield{author}{\bibinfo{person}{Zeping Yu}, \bibinfo{person}{Jianxun Lian}, \bibinfo{person}{Ahmad Mahmoody}, \bibinfo{person}{Gongshen Liu}, {and} \bibinfo{person}{Xing Xie}.} \bibinfo{year}{2019}\natexlab{}.
\newblock \showarticletitle{Adaptive User Modeling with Long and Short-Term Preferences for Personalized Recommendation.}. In \bibinfo{booktitle}{{\em IJCAI}}. \bibinfo{pages}{4213--4219}.
\newblock


\bibitem[\protect\citeauthoryear{Zheng, Gao, Chang, Niu, Song, Jin, and Li}{Zheng et~al\mbox{.}}{2022}]%
        {zheng2022disentangling}
\bibfield{author}{\bibinfo{person}{Yu Zheng}, \bibinfo{person}{Chen Gao}, \bibinfo{person}{Jianxin Chang}, \bibinfo{person}{Yanan Niu}, \bibinfo{person}{Yang Song}, \bibinfo{person}{Depeng Jin}, {and} \bibinfo{person}{Yong Li}.} \bibinfo{year}{2022}\natexlab{}.
\newblock \showarticletitle{Disentangling long and short-term interests for recommendation}. In \bibinfo{booktitle}{{\em Proceedings of the ACM Web Conference 2022}}. \bibinfo{pages}{2256--2267}.
\newblock


\bibitem[\protect\citeauthoryear{Zhou, Mou, Fan, Pi, Bian, Zhou, Zhu, and Gai}{Zhou et~al\mbox{.}}{2019}]%
        {zhou2019deep}
\bibfield{author}{\bibinfo{person}{Guorui Zhou}, \bibinfo{person}{Na Mou}, \bibinfo{person}{Ying Fan}, \bibinfo{person}{Qi Pi}, \bibinfo{person}{Weijie Bian}, \bibinfo{person}{Chang Zhou}, \bibinfo{person}{Xiaoqiang Zhu}, {and} \bibinfo{person}{Kun Gai}.} \bibinfo{year}{2019}\natexlab{}.
\newblock \showarticletitle{Deep interest evolution network for click-through rate prediction}. In \bibinfo{booktitle}{{\em Proceedings of the AAAI conference on artificial intelligence}}, Vol.~\bibinfo{volume}{33}. \bibinfo{pages}{5941--5948}.
\newblock


\bibitem[\protect\citeauthoryear{Zhou, Zhu, Song, Fan, Zhu, Ma, Yan, Jin, Li, and Gai}{Zhou et~al\mbox{.}}{2018}]%
        {zhou2018deep}
\bibfield{author}{\bibinfo{person}{Guorui Zhou}, \bibinfo{person}{Xiaoqiang Zhu}, \bibinfo{person}{Chenru Song}, \bibinfo{person}{Ying Fan}, \bibinfo{person}{Han Zhu}, \bibinfo{person}{Xiao Ma}, \bibinfo{person}{Yanghui Yan}, \bibinfo{person}{Junqi Jin}, \bibinfo{person}{Han Li}, {and} \bibinfo{person}{Kun Gai}.} \bibinfo{year}{2018}\natexlab{}.
\newblock \showarticletitle{Deep interest network for click-through rate prediction}. In \bibinfo{booktitle}{{\em Proceedings of the 24th ACM SIGKDD International Conference on Knowledge Discovery \& Data Mining}}. \bibinfo{pages}{1059--1068}.
\newblock


\end{thebibliography}
\clearpage

\end{document}